\documentclass[twocolumn]{aastex631} 

\usepackage{color}
\usepackage{hyperref}
\usepackage{epstopdf}
\usepackage{graphicx}
\usepackage[FIGTOPCAP]{subfigure}
\usepackage{CJKutf8}
\usepackage{amsmath}
\usepackage{longtable}
\usepackage{threeparttable}

\newcommand{\rj}[1]{\textcolor{black}{#1}}

\begin{document}
\title{{\tt RapidGBM}: An Efficient Tool for Fermi-GBM Visibility Checking and Data Analysis with a Case Study of EP240617a}

\correspondingauthor{Yun Wang, Da-Ming Wei}
\email{wangyun@pmo.ac.cn, dmwei@pmo.ac.cn}

\author[0000-0002-8385-7848]{Yun Wang}
\affiliation{Purple Mountain Observatory, Chinese Academy of Sciences, Nanjing 210023, China}

\author[0000-0002-9037-8642]{Jia Ren}
\affiliation{Purple Mountain Observatory, Chinese Academy of Sciences, Nanjing 210023, China}

\author[0000-0002-2277-9735]{Lu-Yao Jiang}
\affiliation{Purple Mountain Observatory, Chinese Academy of Sciences, Nanjing 210023, China}
\affiliation{School of Astronomy and Space Science, University of Science and Technology of China, Hefei 230026, China}

\author[0000-0003-2915-7434]{Hao Zhou}
\affiliation{Purple Mountain Observatory, Chinese Academy of Sciences, Nanjing 210023, China}

\author[0000-0002-5596-5059]{Yi-Han Iris Yin}
\affiliation{School of Astronomy and Space Science, Nanjing University, Nanjing 210093, China}
\affiliation{Key Laboratory of Modern Astronomy and Astrophysics (Nanjing University), Ministry of Education, China}

\author{Yi-Fang Liang}
\affiliation{Purple Mountain Observatory, Chinese Academy of Sciences, Nanjing 210023, China}
\affiliation{School of Astronomy and Space Science, University of Science and Technology of China, Hefei 230026, China}

\author[0000-0003-4977-9724]{Zhi-Ping Jin}
\affiliation{Purple Mountain Observatory, Chinese Academy of Sciences, Nanjing 210023, China}
\affiliation{School of Astronomy and Space Science, University of Science and Technology of China, Hefei 230026, China}

\author[0000-0002-8966-6911]{Yi-Zhong Fan}
\affiliation{Purple Mountain Observatory, Chinese Academy of Sciences, Nanjing 210023, China}
\affiliation{School of Astronomy and Space Science, University of Science and Technology of China, Hefei 230026, China}

\author[0000-0002-9758-5476]{Da-Ming Wei}
\affiliation{Purple Mountain Observatory, Chinese Academy of Sciences, Nanjing 210023, China}
\affiliation{School of Astronomy and Space Science, University of Science and Technology of China, Hefei 230026, China}

\author{Wei Chen}
\affiliation{National Astronomical Observatories, Chinese Academy of Sciences, 20A Datun Road, Beijing 100101, China}
\affiliation{School of Astronomy and Space Science, University of Chinese Academy of Sciences, 19A Yuquan Road, Beijing 100049, China}

\author{Hui Sun}
\affiliation{National Astronomical Observatories, Chinese Academy of Sciences, 20A Datun Road, Beijing 100101, China}

\author{Jing-Wei Hu}
\affiliation{National Astronomical Observatories, Chinese Academy of Sciences, 20A Datun Road, Beijing 100101, China}

\author{Dong-Yue Li}
\affiliation{National Astronomical Observatories, Chinese Academy of Sciences, 20A Datun Road, Beijing 100101, China}

\author[0000-0002-5596-5059]{Jun Yang}
\affiliation{Institute for Astrophysics, School of Physics, Zhengzhou University, Zhengzhou 450001, People’s Republic of China}

\author{Wen-Da Zhang}
\affiliation{National Astronomical Observatories, Chinese Academy of Sciences, 20A Datun Road, Beijing 100101, China}

\author{Yuan Liu}
\affiliation{National Astronomical Observatories, Chinese Academy of Sciences, 20A Datun Road, Beijing 100101, China}

\author{Wei-Min Yuan}
\affiliation{National Astronomical Observatories, Chinese Academy of Sciences, 20A Datun Road, Beijing 100101, China}
\affiliation{School of Astronomy and Space Science, University of Chinese Academy of Sciences, 19A Yuquan Road, Beijing 100049, China}

\author[0000-0002-6299-1263]{Xue-Feng Wu}
\affiliation{Purple Mountain Observatory, Chinese Academy of Sciences, Nanjing 210023, China}
\affiliation{School of Astronomy and Space Science, University of Science and Technology of China, Hefei 230026, China}

\begin{abstract}
We have developed a lightweight tool, {\tt RapidGBM}, featuring a web-based interface and capabilities of rapid calculation of Fermi Gamma-ray Burst Monitor (GBM) visibilities and performance of basic data analysis. It has two key features: (1) it can immediately check the visibility of Fermi-GBM for new transients, and (2) it can check the light curve and perform spectral analysis after the hourly Time-Tagger Event data are released. The visibility check and the response matrix generation required for spectral analysis can be achieved through the historical pointing file after the orbit calculation, even when the real-time pointing file is not yet available. As a case study, we apply the tool to EP240617a, an X-ray transient triggered by Einstein Probe (EP). We demonstrate the workflow of visibility checking, data processing, and spectral analysis for this event. The results suggest that EP240617a can be classified as an X-ray-rich gamma-ray burst (XRR) and confirm the feasibility of using historical pointing files for rapid analysis. Further, we discuss possible physical interpretations of such events, including implications for jet launching and progenitor scenarios. Therefore, {\tt RapidGBM} is expected to assist EP Transient Advocates, Space-based multiband astronomical Variable Objects Monitor burst advocates, and other members of the community in cross checking high-energy transients. Based on prompt emission parameter relations (e.g. $E_{\rm p}$-$E_{\gamma,\rm iso}$), it can also help identify peculiar GRBs (e.g. long-short burst, magnetar giant flare, etc.) and provide useful references (e.g. more accurate $T_0$) for scheduling follow-up observations.
\end{abstract}

\section{Introduction} \label{sec:intro}
Gamma-ray bursts (GRBs) are among the most energetic and enigmatic transients in the Universe, providing unique laboratories for studying extreme physics, relativistic jets, and the deaths of massive stars \citep{1993ApJ...405..273W,2006Natur.441..463F} or compact object mergers \citep{1996bboe.book..682E,1992ApJ...395L..83N,2005Natur.437..851G,2010ApJ...708....9F,2010ApJ...725.1202L,2013ApJ...776...18F,2014ARA&A..52...43B}. With the successful launch and scientific operation of the Einstein Probe \citep[EP,][]{2025arXiv250107362Y}, several GRBs have already been detected to date, including EP240219a/GRB 240219A \citep{2024ApJ...975L..27Y}, EP240315a/GRB 240315C \citep{2024arXiv240416350L,2025NatAs...9..564L}, EP240801a/XRF 240801B \citep{,2025arXiv250304306J}, and GRB 250404A/EP250404a \citep{2025arXiv250600435Y}. These achievements are enabled by EP’s wide-field soft X-ray monitoring and rapid follow-up capabilities. The Wide-field X-ray Telescope (WXT) onboard EP, operating in the 0.4–5 keV band, provides an instantaneous field of view of approximately 3600 deg$^{2}$, making it highly efficient for transient detection. Complementing this, the Follow-up X-ray Telescope (FXT), covering the 0.3–10 keV range with a larger effective area, enables prompt follow-up observations with improved localization accuracy. The FXT can trace the early temporal evolution of X-ray afterglows with a typical sensitivity reaching $\sim10^{-14}$ erg/cm$^{2}$/s. The coordinated observations of WXT and FXT allow EP to detect softer GRBs than previous missions and to monitor their early-time light curves and spectral evolution in the X-ray band, which are critical for probing the physical origin of GRBs.

Meanwhile, the Fermi Gamma-ray Burst Monitor \citep[GBM,][]{meegan2009fermi} continues to serve as an indispensable instrument for GRB detection, providing continuous all-sky coverage in the $\gamma$-ray band. For each new GRB triggers, especially those discovered by EP, rapid assessment of Fermi-GBM visibility, data availability, and preliminary spectral analysis are critical for the rapid evaluation of scientific values and coordinating follow-up observations. Duty Advocates require fast and reliable tools to determine whether a GRB location was within the GBM field of view at the time of trigger, to check for the presence of significant signals, and to perform preliminary analysis and coordinate follow-up observations. This is particularly important for triggers from WXT, which may correspond to GRBs below the onboard threshold of GBM.

To address these requirements, we developed {\tt RapidGBM}\footnote{\url{https://github.com/0neyun/RapidGBM}}: an efficient, interactive web application designed specifically for the rapid checking of Fermi-GBM visibility and data analysis \citep{RapidGBM}. {\tt RapidGBM} allows users to instantly check GBM coverage for any sky position and time, automates the generation of detector response files, and supports quick-look light curve and spectral analysis. The tool streamlines GBM data analysis workflows, assisting duty teams of EP in rapid event response, especially for sub-threshold GBM events.

In this paper, we first introduce the design and capabilities of {\tt RapidGBM} in Section 2. In Section 3, we present a case study of EP240617a, demonstrating the application of the tool to data analysis during its prompt emission phase. In Section 4, we summarize and discuss the {\tt RapidGBM} tool and EP240617a in detail, including the possible physical explanation of this burst and the potential use of this tool in future GRB follow-up observations.

\section{Design and Capabilities of \texttt{RapidGBM}} \label{sec:rapidgbm}
\texttt{RapidGBM} is a lightweight, web-based tool designed to support the quicklook analysis of transient events with respect to Fermi-GBM coverage and data products. The tool use public {\tt Fermi GBM tools} \citep{GbmDataTools} APIs, and the orbital calculation draw upon the {\tt osv}\footnote{\url{https://fermi.gsfc.nasa.gov/ssc/data/analysis/user/Fermi_GBM_OrbitalBackgroundTool.pdf}} \citep{2012SPIE.8443E..3BF} from Fermi User Contributions. Response Matrix Files are generated with the {\tt GBM Response Generator}\footnote{\url{https://fermi.gsfc.nasa.gov/ssc/data/analysis/gbm/DOCUMENTATION.html}} provided by the Fermi team. Spectral fitting uses the {\tt PyXspec}\footnote{\url{https://heasarc.gsfc.nasa.gov/xanadu/xspec/python/html/index.html}} package. The interactive web interface is built with the easily deployable {\tt streamlit}\footnote{\url{https://streamlit.io}} framework, making it portable and easy to deploy across different platforms. Its intuitive interface allows users to complete a full visibility check and spectral analysis within minutes after data become available. The tool has already been tested and used during duty shifts by both EP Transient Advocates (EP-TAs) and SVOM Burst Advocates (SVOM-BAs). It consists of three main functional components, as described below.

\subsection{GBM Visibility Checker} 
Given a sky coordinate and a specific time, \texttt{RapidGBM} determines whether the source is observable by Fermi-GBM or obscured by the Earth, or whether the spacecraft is within the South Atlantic Anomaly (SAA). The calculation is based on the position history (poshist) file. When both the spacecraft position and the time are specified, the tool calls relevant APIs from the {\tt Fermi GBM tools} to compute visibility and generate diagnostic plots, as shown in the examples below. 

We can use either the real-time poshist file or the historical poshist file. When using a historical poshist file, the key step is to determine the appropriate reference time that best matches the spacecraft’s orbital state at the time of the event. Motivated by the method of orbital background estimation \citep{2012SPIE.8443E..3BF}, we use times when the satellite has the same geographical footprint to approximate the visibility at the time of interest. In practice, we adopt the data from 30 orbits prior to the external trigger (EP or SVOM) time, {assuming that the Fermi spacecraft has not executed any pointing maneuver or change in its rocking profile, as it returns to nearly the same geographical coordinates every 15 orbits and the detector pointing geometry repeats every 2 orbits (for $|\beta| < 24^\circ$, where $\beta$ is the angle between the orbital plane and the Sun).} Previous studies have shown that using data from $\pm$30 orbits provides a reliable estimate of the spacecraft environment, particularly for evaluating visibility and SAA passage conditions \citep{2012SPIE.8443E..3BF}. Assuming that the Fermi spacecraft follows an approximately circular orbit, its orbital period can be estimated from the average geocentric distance using Newtonian gravity. Following the method described in the {\tt osv} tool (v1.3), the orbital period $T$ is given by:
\begin{equation}
T = 2\pi \sqrt{\frac{r_{\mathrm{avg}}^3}{G M_{\oplus}}}
\label{eq:orbital_period}
\end{equation}
where $T$ is the orbital period in seconds, $r_{\mathrm{avg}}$ is the average orbital radius (in meters), $G = 6.67428 \times 10^{-11}\ \mathrm{m^3\,kg^{-1}\,s^{-2}}$ is the gravitational constant, $M_{\oplus} = 5.9722 \times 10^{24}\ \mathrm{kg}$ is the mass of the Earth. The average orbital radius $r_{\mathrm{avg}}$ is computed from the position vectors of the spacecraft:
\begin{equation}
r_{\mathrm{avg}} = \frac{1}{N} \sum_{i=1}^N \sqrt{x_i^2 + y_i^2 + z_i^2}
\label{eq:average_radius}
\end{equation}
where $(x_i, y_i, z_i)$ are the Cartesian coordinates of the spacecraft at time index $i$, and $N$ is the total number of position samples. All the necessary parameters are provided in the poshist file. Using data from the previous two days, we compute the time corresponding to 30 orbits prior to the external trigger time and use it to check the GBM visibility at that reference time. This approach is particularly useful when the current day’s poshist file is not yet available, or when evaluating sub-threshold GRB candidates that did not trigger onboard, to determine whether the location of the external trigger was within the GBM field of view.

\subsection{TTE Data Handler and Response Generator}
Once continuous Time-Tagged Event (TTE) data are released (typically within a few hours), \texttt{RapidGBM} automatically checks for and downloads the corresponding data. The tool then calls relevant APIs from the {\tt Fermi GBM tools} to generate light curves, perform background fitting and subtraction, and compute the signal-to-noise ratio \citep[SNR;][]{1983ApJ...272..317L} to assess the presence of potential associated signals. Users can extract both source and background spectra over user-defined time intervals. Background estimation is typically based on pre- and postburst intervals selected through the interface. Detector response files are generated using the {\tt GBM Response Generator}. After determining the orbital period during the visibility check, the tool can also use historical {poshist} and {cspec} files to generate appropriate response matrices when the current day’s data are not yet available. 

\begin{figure}[]
    \centering
    \includegraphics[width=0.45\textwidth]{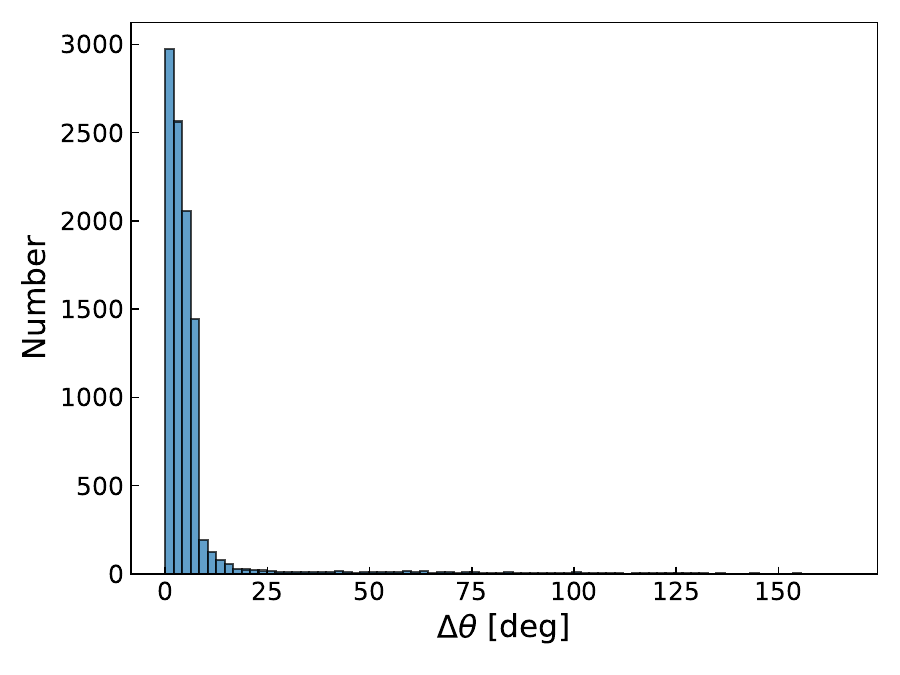}
    \caption{Distribution of angular deviations obtained by simulating the use of historical poshist files.}
    \label{fig:delta_angle}
\end{figure}

{It is worth noting that generating response files based on historical pointing carries certain risks. Although in early 2018 the Fermi spacecraft experienced an anomaly with its -Y Solar Array Drive Assembly (SADA), after which Autonomous Repoint (ARR) and Target of Opportunity (ToO) observations have essentially no longer been carried out, Fermi currently operates with a mixture of rocking profiles that can still alter its pointing. To evaluate the reliability of using historical pointing, we performed a simple simulation in which 10,000 random coordinates and times (between 2019-01-01 and 2025-01-01 UTC) were generated to examine the angular deviations. The results indicate a mean deviation of $6.7^\circ$, with $\sim$ 8\% of cases exceeding $10^\circ$, and the distribution of deviation angles ($\Delta \theta$) is shown in Figure~\ref{fig:delta_angle}. Therefore, we recommend caution when using historical pointing files to generate response files and advise verifying the Observatory Status online\footnote{\url{https://fermi.gsfc.nasa.gov/ssc/observations/timeline/posting/}} to determine whether the spacecraft has changed its rocking profile or executed any other pointing maneuvers. Moreover, spectral fitting results based on response files generated from historical pointing should be regarded as preliminary.}

This module generates all the files required for GBM spectral analysis.

\subsection{Spectral Analysis Module} 
Spectral fitting is performed with the {\tt PyXspec} package, which provides access to the standard models in {\tt XSPEC}\footnote{\url{https://heasarc.gsfc.nasa.gov/docs/xanadu/xspec/index.html}}. The results can include key spectral parameters such as the photon index, the peak energy ($E_{\rm peak}$), and the energy flux, together with model statistics and confidence intervals. This module is designed to enable rapid assessment of the spectral properties of transient sources, which is particularly useful for identifying GRB-like signatures in untriggered events. Its lightweight interface allows community members to carry out preliminary spectral diagnostics within minutes after the hourly TTE data become available, and it also supports uploading any files readable by {\tt XSPEC} for joint spectral analysis.

\section{Case Study: EP240617{\lowercase{a}}} \label{sec:Obs}
\subsection{Observations}
\begin{figure}[]
    \centering
    \includegraphics[width=0.49\textwidth]{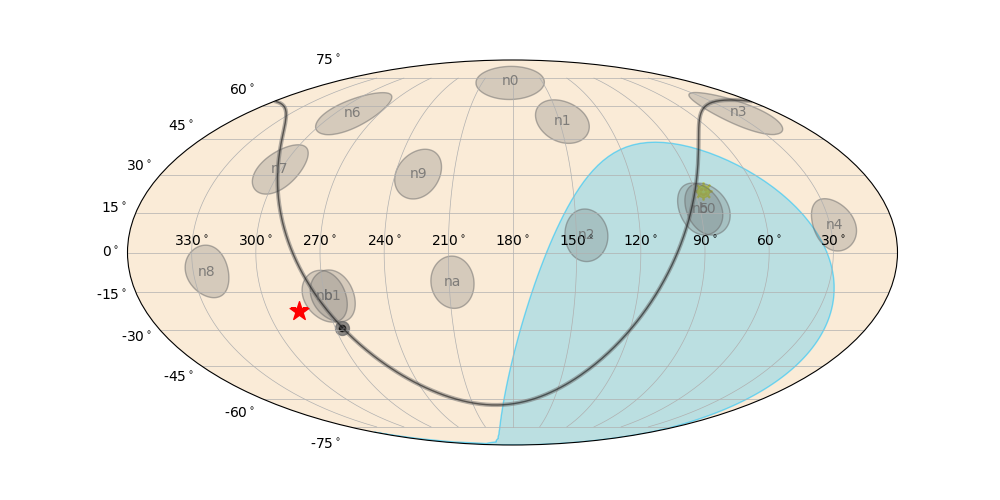}
    \includegraphics[width=0.49\textwidth]{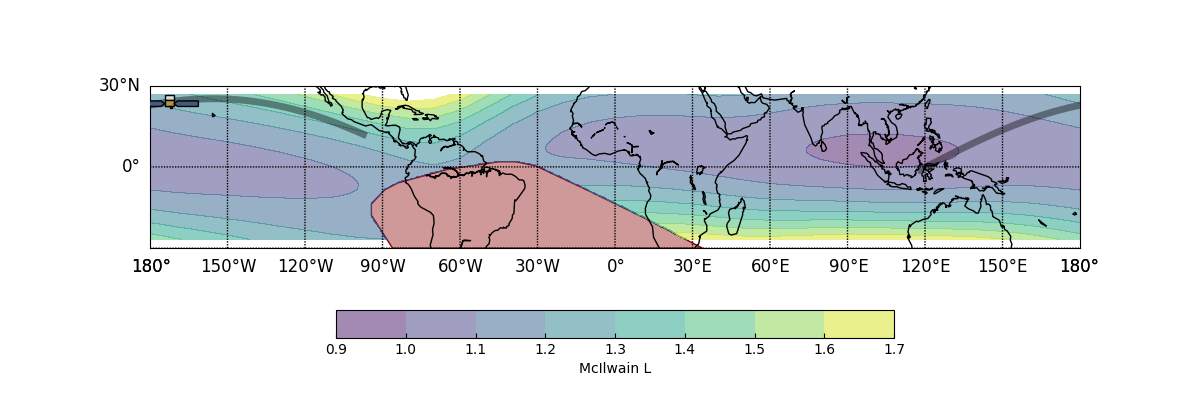}
    \caption{GBM visibility map at the time of EP240617a. In the upper panel, the blue area indicates the region obscured by the Earth, while the red pentagram marks the source position. In the lower panel, the red area represents the SAA region, and the satellite icon traces the orbit 20 minutes before and after the event.}
    \label{fig:gbm_visibility}
\end{figure}

EP-WXT detected an X-ray transient source EP240617a, which started at 12:19:13 UTC on June 17, 2024, and the position of the source is R.A.~=~285.030 deg, decl.~=~-22.561 deg (J2000) with an uncertainty of 3 arcmin in radius (90\% C.L. statistical and systematic) \citep{GCN36691}. We examined the visibility of this position by the Fermi-GBM detectors at the time of the event. As shown in Figure~\ref{fig:gbm_visibility}, the source was not occulted by the Earth. The nearest GBM NaI detector is nb (15.1$^\circ$ ). When using the historical pointing files, the calculated angle is $15.6^\circ$, which is in good agreement with the result obtained using the real-time pointing files. In addition, the spacecraft was not scheduled to enter or exit the South Atlantic Anomaly (SAA) within $\pm$20 minutes of the event time.

\begin{figure}[]
    \centering
    \includegraphics[width=0.45\textwidth]{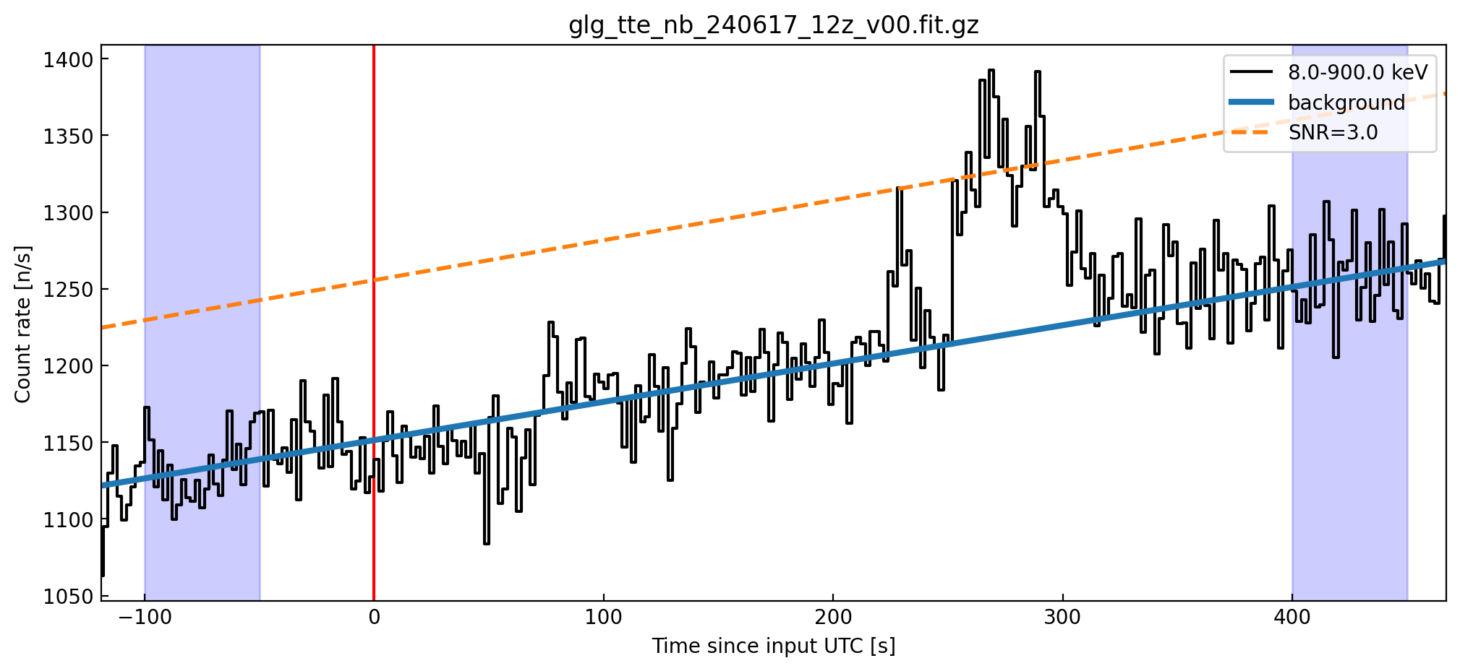}
    \includegraphics[width=0.45\textwidth]{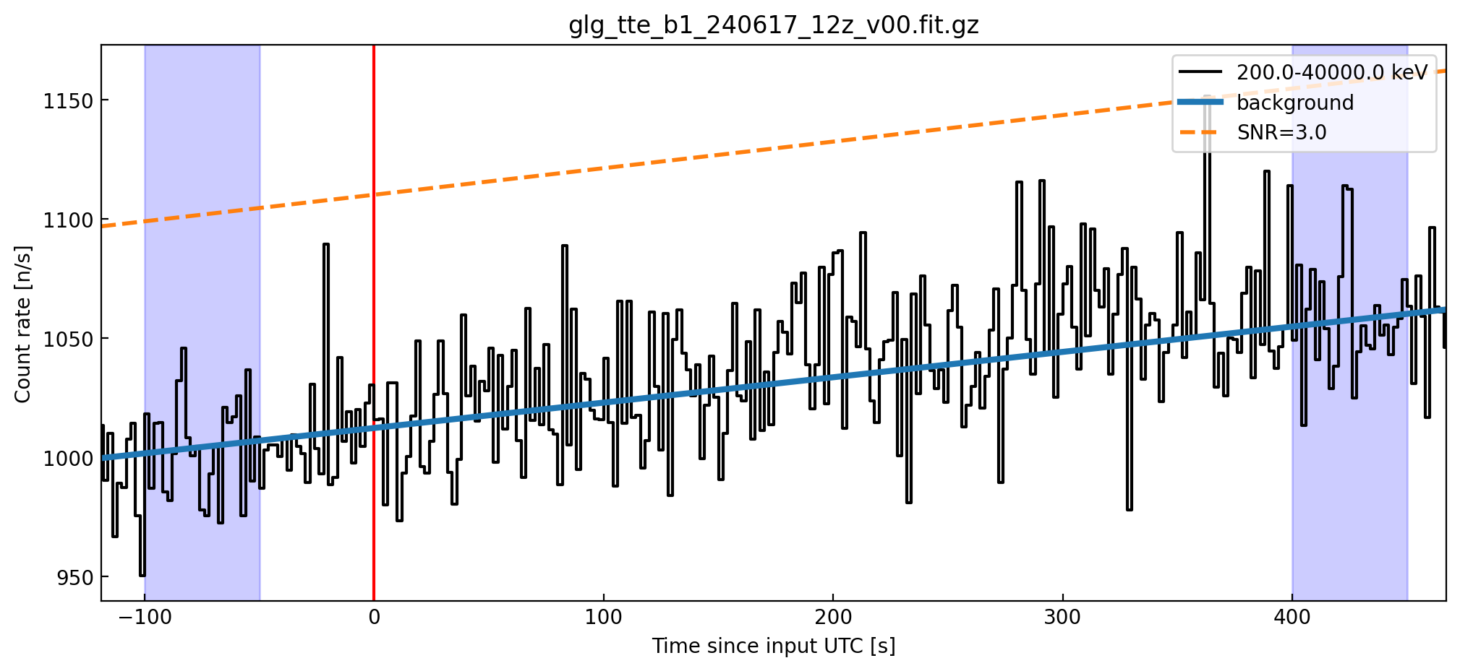}
    \caption{Light curve of EP240617a generated from nb and b1 detector TTE data. The blue shaded area indicates the interval used for background fitting, the blue and orange solid lines are the fitted background and the count rate corresponding to SNR = 3, respectively, and the red solid line is the trigger time of EP240617a.}
    \label{fig:gbm_lc}
\end{figure}

We used the TTE data from the \texttt{nb} detector to generate the light curve of EP240617a from $T_0 - 120$ s to $T_0 + 470$ s, as shown in the upper panel of Figure~\ref{fig:gbm_lc}. A significant excess signal was found in the 8–900 keV energy range, with a SNR greater than 3, consistent with the Circular report \citep{2024GCN.36692....1Y}. The lower panel of Figure~\ref{fig:gbm_lc} shows the light curve of the \texttt{b1} detector in the 200–40,000 keV energy range, where no significant signal was detected. Since most of the photons are below 200 keV, in the following analysis we focus only on the NaI (nb) data. Therefore, this event is likely a sub-threshold GRB that did not trigger the GBM and is not included in the official GBM trigger catalog.

Unfortunately, the follow-up observations were limited, with only a few Circulars reporting optical upper limits \citep{2024GCN.36693....1P,2024GCN.36707....1S} and a possible X-ray afterglow \citep{2024GCN.36722....1S}. In the following, we will primarily focus on the early prompt emission of this event, specifically the observations from EP-WXT and Fermi-GBM.

\subsection{Data reduction of EP-WXT and Fermi-GBM}
\begin{figure}[!htp]
    \centering
    \includegraphics[width=0.45\textwidth]{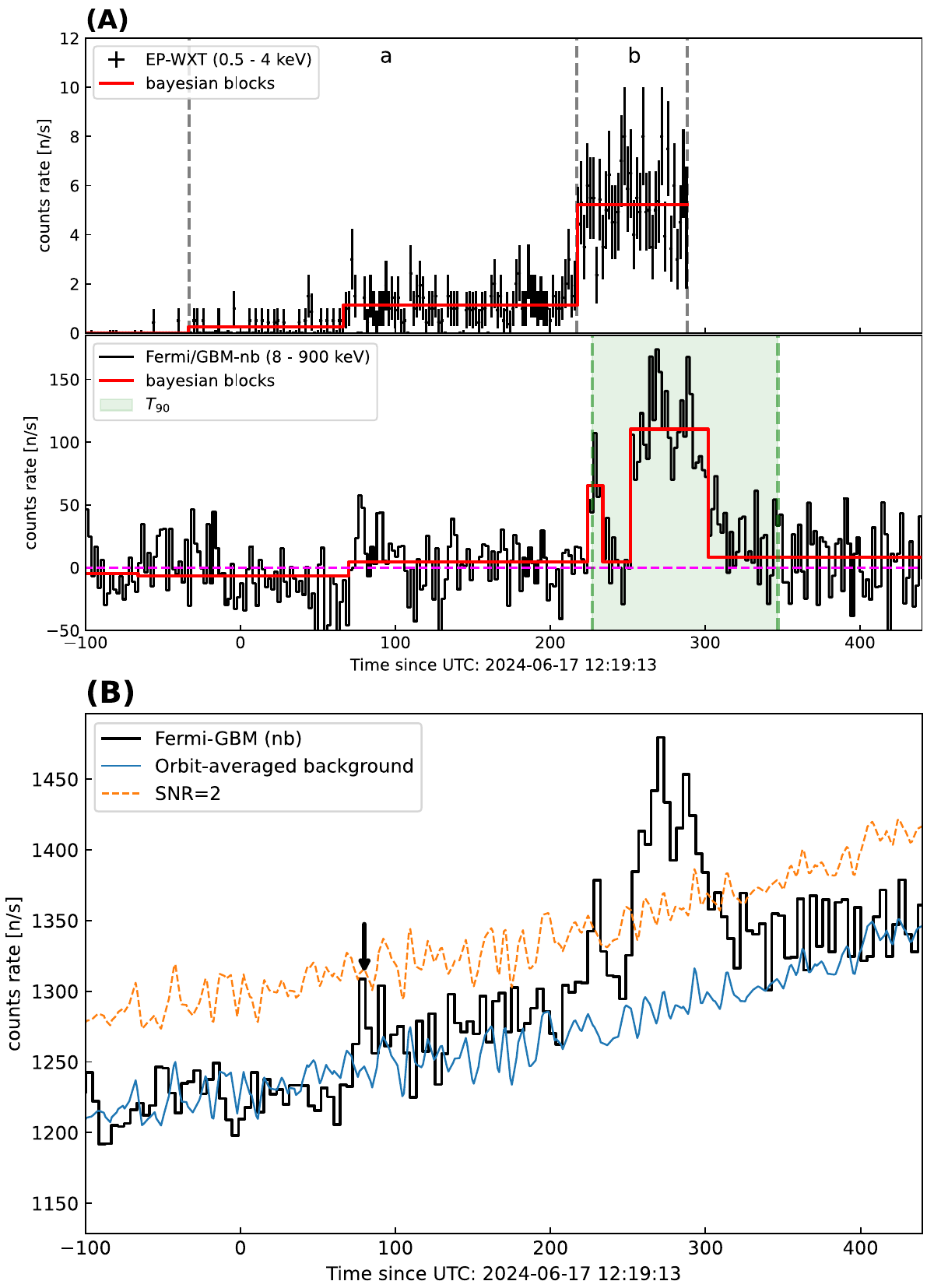}
    \caption{The light curve of EP240617a. Panel (A) shows the background-subtracted light curves for WXT and GBM, with the red solid line indicating the Bayesian block result. The green-shaded region indicates the $T_{90}$ interval of the GBM data. Panel (B) shows the GBM light curve without background subtraction. The blue curve represents the orbit-averaged background, while the orange line indicates the background-derived count rate corresponding to SNR = 2. The black arrow indicates a time around 80 seconds, at which a corresponding increase is also observed in the WXT data.}
    \label{fig:WXT_GBM_LC}
\end{figure}

The WXT data were reduced using the standard procedures in the WXT Data Analysis Software ({\tt WXTDAS}, v2.10; Liu et al., in preparation) using the latest calibration database \citep[CALDB;][]{Cheng2025}, and the pipeline was employed to generate images, light curves, and spectra. The upper panel of Figure~\ref{fig:WXT_GBM_LC} (A) shows the background-subtracted light curve of WXT in the 0.5–4 keV energy band. We analyzed this light curve using the Bayesian blocks algorithm \citep{scargle2013studies} and found that the signal began approximately 33 s before the trigger time (2024-06-17T12:19:13) as reported in \citet{GCN36691}. The light curve clearly exhibits two distinct phases:  a “smooth” emission phase from $T_0 - 33$ s to $T_0 + 217$ s (a) and a “main burst” phase from $T_0 + 217$ s to $T_0 + 289$ s (b), coinciding with the gamma-ray activity.
For the continuous TTE data from the Fermi-GBM {nb} detector, the background-subtracted light curve in the 8–900 keV energy range is shown in the lower panel of Figure~\ref{fig:WXT_GBM_LC} (A). We also analyzed this light curve using the Bayesian blocks algorithm and identified two distinct signals. Additionally, by calculating the cumulative photon counts \citep{koshut1996systematic}, we determined a $T_{90}$ duration of is 120 $\pm$ 4 s, spanning from $T_0 + 227$ s to $T_0 + 347$ s. We used {\tt RapidGBM} to extract spectra for two time intervals for spectral analysis: one corresponding to the time interval of the WXT (b) phase, and the other covering the $T_{90}$ interval.

As shown in panel (B) of Figure~\ref{fig:WXT_GBM_LC}, in addition to fitting the background by selecting intervals before and after the burst, we also considered using the orbit-averaged background, a method commonly used in background testing and source detection. This approach uses detection rates from adjacent days, when the satellite is at the same geographical coordinates, to estimate the background at the time of interest~\citep{2011arXiv1111.3779F,2012SPIE.8443E..3BF}. Compared with polynomial background subtraction, the orbital subtraction technique is more suitable for analyzing long-lived emission and is more effective at identifying weak emission signals~\citep{2014ApJ...789...20A,2015MNRAS.452..824K,2023ApJ...952L..42L}. The Fermi satellite was at the same coordinates every $\sim 15$ orbits (corresponding to $\sim 24$ hr), and the spacecraft rocking angle was the same every two orbits. Thus, we use the average of data from $\pm30$ orbits to estimate the background of EP240617a, as shown by the blue line in panel (B) of Figure~\ref{fig:WXT_GBM_LC}. It is evident that around 80 seconds after $T_0$, a weak signal appears in the GBM data (indicated by the black arrow), with a signal strength close to the count rate corresponding to SNR = 2, consistent with the flux variations observed by WXT at the same time. 

\subsection{Spectral analysis}
We separately performed spectral fits for the WXT and GBM data in the above time intervals, as well as joint fits that combine both datasets. The photon spectral models considered in the fits were a simple power-law (PL) model and a cutoff power-law (CPL) model, expressed as:

\begin{equation}
N(E) = K E^{-\Gamma}
\end{equation}

and

\begin{equation}
N(E) = K E^{-\Gamma} \exp\left(-\frac{E}{E_{\rm peak}/(2-\Gamma)}\right),
\end{equation}

where $N(E)$ is the photon flux at energy $E$, $K$ is the normalization constant, $\Gamma$ is the photon index, and $E_{\rm peak}$ is the peak energy in $\nu F\nu$ spectrum. In addition, we used the {$tbabs$} model to account for the soft X-ray absorption along the line of sight. Since the distance to this burst is unknown, we set the Galactic hydrogen column density ($N_{\rm H}$ = 0.169 $\times$ 10$^{22}$ cm$^{-2}$) as the lower limit for the absorption model in our fits. We used the online {\tt NHtot tool}\footnote{\url{https://www.swift.ac.uk/analysis/nhtot/index.php}} provided on the Swift website to calculate $N_{\rm H}$. For full details, see \cite{2013MNRAS.431..394W}.

\begin{figure}[!htp]
    \centering
    \includegraphics[width=0.49\textwidth]{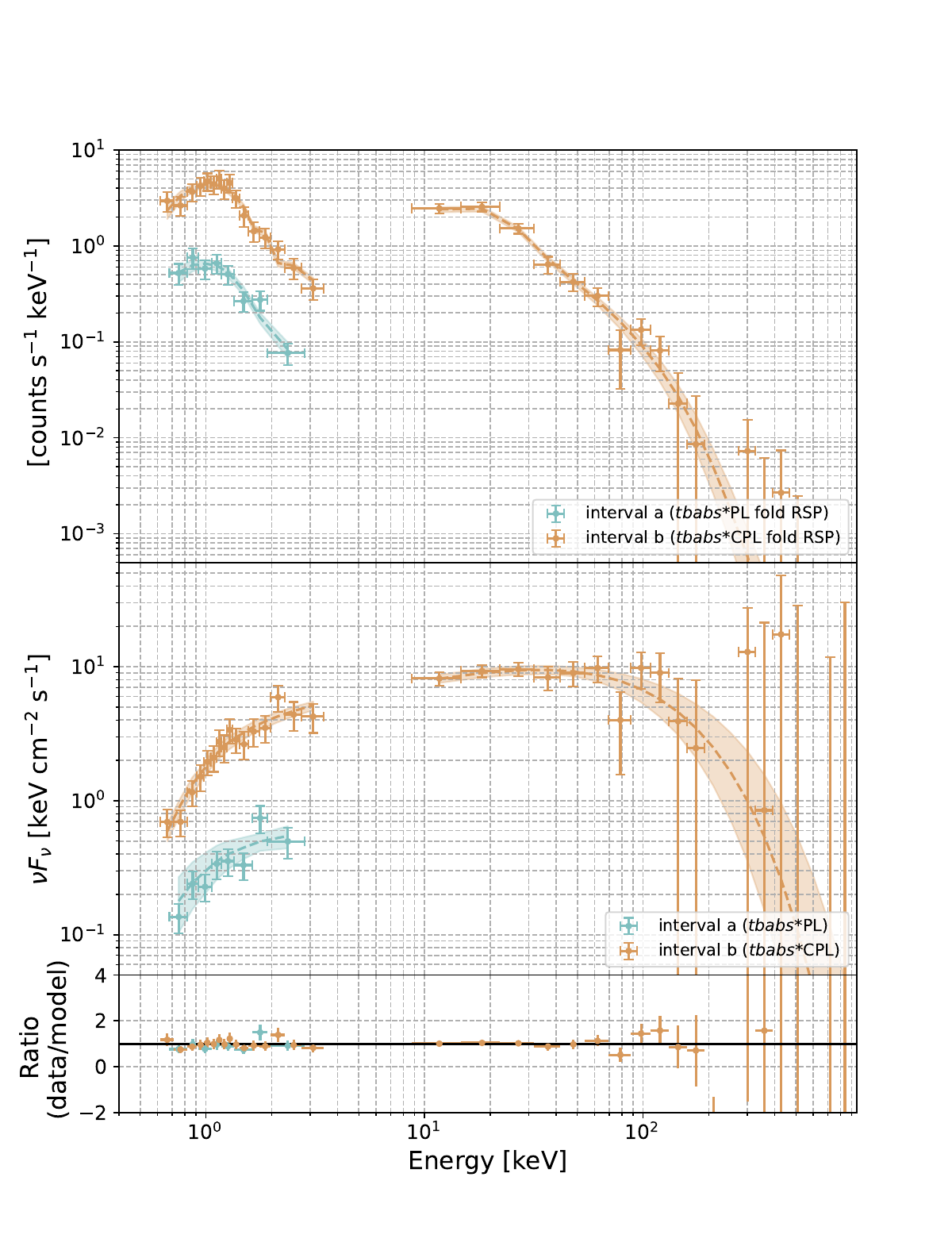}
    \caption{{The spectral analysis of EP240617a for intervals a and b. The top panel shows the observed count spectra with best-fit models, with shaded regions indicating the 90\% confidence intervals. The middle panel presents the corresponding $\nu F_\nu$ spectra with the same confidence regions, and the bottom panel displays the data-to-model ratios, with interval a in light blue and interval b in light orange.}}
    \label{fig:eeuf}
\end{figure}

In the spectral fitting, we used {\tt PyXspec} to perform the forward-folding convolution of the photon spectra with the instrument response matrices and to calculate the statistics, and we employed {\tt PyMultiNest} as the sampler for Bayesian inference. For the WXT data, the statistic used was the Cash statistic ({\tt cstat}), while for the GBM data, the statistic used was the Poisson Gaussian statistic ({\tt pgstat}), as detailed in the {\tt XSPEC} manual\footnote{\url{https:// heasarc.gsfc.nasa.gov/xanadu/xspec/manual/XSappendixStatistics.html}}. Spectral fits were performed in the 0.5–4 keV band for WXT and in the 8–900 keV range for GBM. We also performed model selection in our fitting analysis using Bayesian evidence as the selection criterion. Typically, a natural logarithm of the Bayes factor, $\log \rm{BF}^{A}_{B} \equiv \log (\mathcal{Z}_{A}) - \log (\mathcal{Z}_{B})$, greater than 8 is considered to indicate strong support for model A over model B \citep{Thrane2019,jeffreys1998theory}. 

As shown in Table \ref{tab:specfit}, we present the results of the spectral fits for WXT and GBM in different time intervals, as well as their joint fits. For interval b, we performed joint fits using both the $tbabs$*PL model and the $tbabs$*CPL model. The inclusion of an additional peak energy component in the latter model leads to a slight improvement in the fit, with the natural logarithm of the Bayes factor (lnBF) equal to 7.23 (see Table \ref{tab:specfit}). Figure \ref{fig:eeuf} shows the photon count spectra, the $\nu F_\nu$ spectra, and the ratios of model to data for the two time intervals.

In addition, we compared the differences between the response matrices generated using the historical pointing files and those generated using the real-time pointing files during the fitting process. {As shown in Figure \ref{fig:spec_compare}, for this event the results are generally consistent within the error margins, provided that the deviation between the historical and actual pointing is minimal ($\sim$ 0.5°), suggesting that this approach may be useful for rapid early-phase analysis under small pointing deviations.}

\subsection{Characteristics}
\begin{figure}[!htp]
    \centering
    \includegraphics[width=0.49\textwidth]{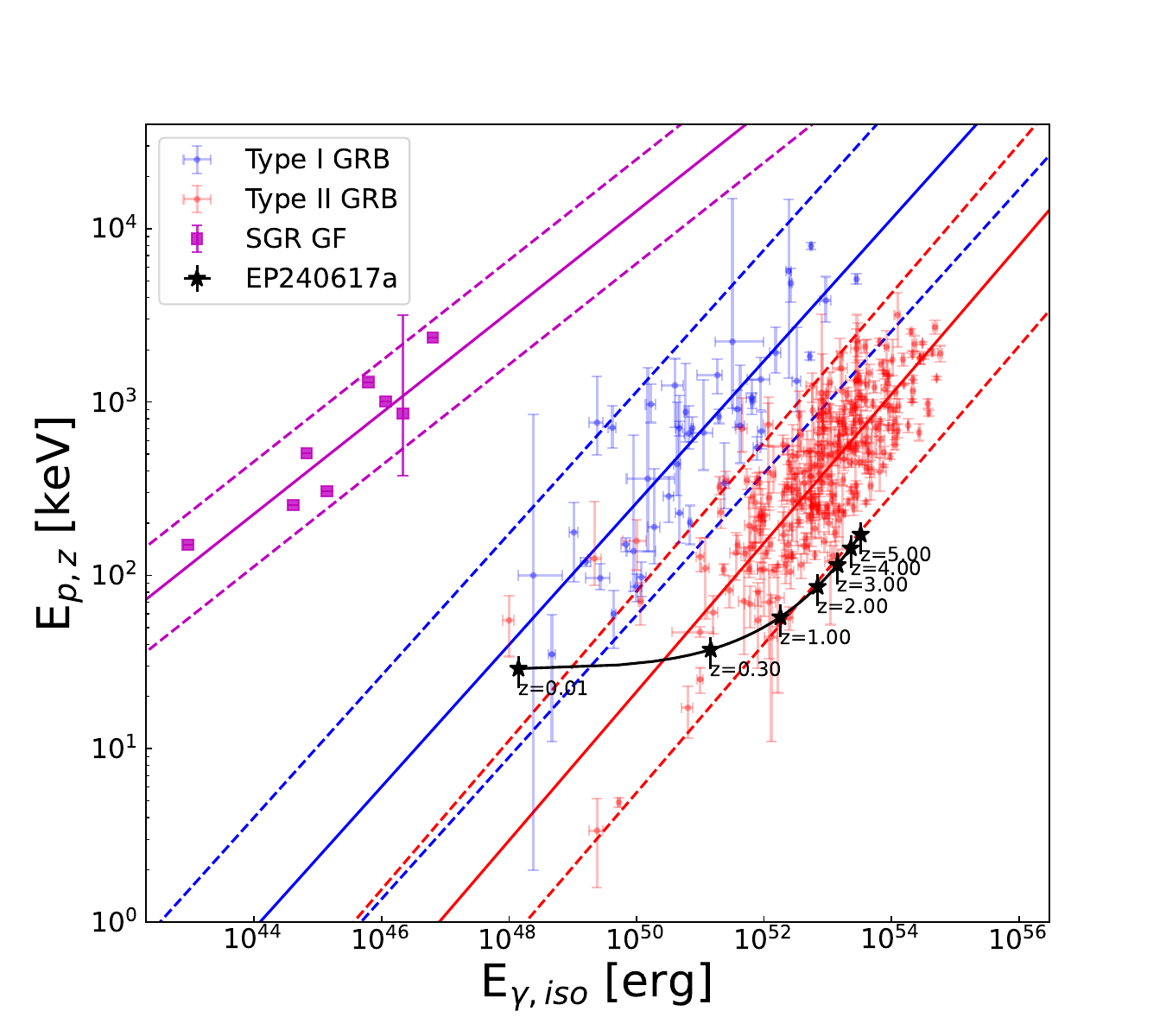}
    \caption{The $E_{\rm p,z}$–$E_{\gamma,\rm iso}$ diagram. The light blue, and light red, and purple points represent the data of Type I and Type II GRBs, and Soft Gamma-ray Repeater Giant Flares (SGR GFs), respectively, with known distances \citep{2006Natur.444.1010Z,2009ApJ...703.1696Z,minaev2020p,2020AstL...46..573M}. The corresponding dashed lines indicate the 2$\sigma_{\rm cor}$ confidence regions of the correlations. The black pentagrams represent the calculated results for EP240617a at different redshifts, while the black curve indicates the results for redshifts varying from 0.01 to 5.}
    \label{fig:amati}
\end{figure}

Considering that the WXT coverage of the source was incomplete (as the source moved out of the field of view), we used the fit results from the $T_{90}$ interval of the GBM data to calculate the $\gamma$-ray fluence, obtaining $S_\gamma = 6.29_{-0.57}^{+0.60} \times 10^{-6}\rm{erg/cm^{2}}$ in the 1–10,000 keV band. Based on
\begin{equation}
E_{\gamma,\text{iso}} = \frac{4 \pi d_L^2 k S_{\gamma}}{1+z},
\label{eq:E_gamma_iso}
\end{equation}
where $d_L$ is the luminosity distance and $k \equiv \int^{10^4/(1+z)}_{1/(1+z)} E N(E) dE / \int^{e_2}_{e_1} E N(E) dE$ (with $e_1$ and $e_2$ denoting the detector’s energy band) is the $k$-correction factor \citep{2001AJ....121.2879B}. Assuming different redshifts (z=0.01–5), the relation between the rest-frame peak energy $E_{\rm p,z}$ and the isotropic $\gamma$-ray energy $E_{\gamma,\text{iso}}$ \citep[i.e., the Amati relation;][]{amati2002intrinsic} is shown in Figure \ref{fig:amati}. 
{In the context of this relation, EP240617a would fall in the Type~I GRB region if located at a close distance; however, its long-duration profile shares features commonly observed in Type~II GRBs. It should be noted that the Amati relation has not strictly accounted for off-axis effects, and the origin of the progenitor cannot be truly clarified, so we provide an extensive discussion in the next section.
}

\begin{figure*}
    \centering
    \includegraphics[width=0.85\textwidth]{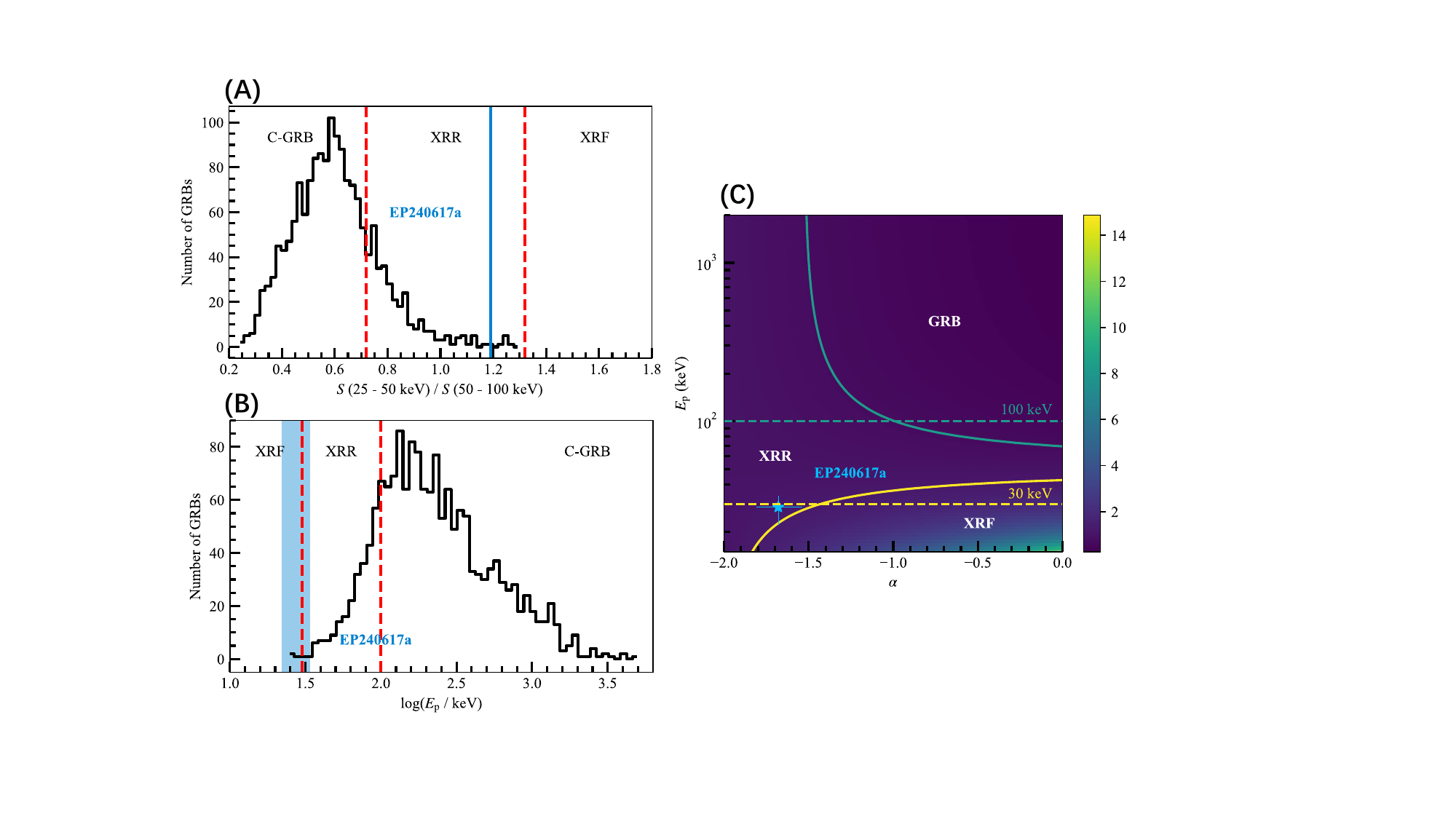}
    \caption{Comparison of EP240617a with historical XRRs and XRFs. Panels (A) and (B) show the distributions of fluence ratios and $E_{\rm p}$, respectively, for Fermi-GBM detected GRB samples. The red dashed vertical lines mark the subclass boundaries of GRBs and are labeled accordingly. The blue vertical lines and shadow area indicate the position of EP240617a within each distribution. Panel (C) shows the CPL parameter space, with photon index and $E_{\rm p}$ on the x- and y-axes, respectively. The color scale represents the fluence ratio $S(25$–$50~\mathrm{keV})/S(50$–$100~\mathrm{keV})$. Solid and dashed lines indicate subclassification boundaries based on fluence ratio and $E_{\rm p}$. EP240617a is marked by a blue star with 1$\sigma$ error bars.
    }
    \label{fig:XRR}
\end{figure*}

As shown in Figure \ref{fig:XRR}, to compare with known GRBs that exhibit more dominant X-ray emission, such as X-ray flashes (XRFs) and X-ray-rich GRBs (XRRs) \citep{2001grba.conf...16H,2003AIPC..662..244K,2005A&A...440..809B,2003A&A...400.1021B,2004ASPC..312...12A,2007A&A...461..485S}, and following the approach of \cite{2024ApJ...975L..27Y}, we also calculated the fluence ratios in different energy bands, such as $S(25$–$50~\rm{keV})$/$S(50$–$100~\rm{keV})$ \citep{2008ApJ...679..570S}, yielding values of 1.19. 
Combined with the observed $E_{\rm p} = 28.68_{-6.63}^{+5.22}~\rm{keV}$ and photon index of $1.68_{-0.14}^{+0.13}$, it may represent an event intermediate between XRR and XRF.

\section{Summary and Discussion}\label{sec:sd}
In this work, we provide a summary from two perspectives. First, we presented {\tt RapidGBM}, a lightweight and efficient tool for performing Fermi-GBM visibility checks and preliminary data analysis. This tool make use of historical pointing files for orbital calculations, enabling key features such as: 
\begin{itemize}
    \item Instant visibility determination of GBM at the trigger time.
    \item Immediate spectral analysis once the hourly TTE data are released.
    \item Consistent spectral fitting results obtained using response matrices generated from historical pointing files compared to those using real-time pointing files.
\end{itemize}
We also note that the Fermi official website provides a publicly available tool for signal searches, {\tt gamma-ray-targeted-search}\footnote{\url{https://github.com/USRA-STI/gamma-ray-targeted-search}}, which performs a more advanced analysis by accounting for both direct detector responses and atmospheric scattering effects. While our signal identification relies on orbit-averaged data and polynomial background fitting, the official tool offers a more rigorous approach to detection. Our primary goal, however, is to provide an interactive platform that enables users such as EP-TAs or SVOM-BAs to quickly examine data and support follow-up planning. {\tt RapidGBM} is one of the modules within the {\tt HEtools} \citep{2023ApJ...953L...8W} framework. In addition, the interactive web interface\footnote{\url{https://hetools.xyz}} offers other tools to support duty scientists, including an upper-limit calculator, a GCN digester powered by large language models, and an afterglow modeling module.

Second, we applied this tool to the analysis of EP240617a, with its main characteristics summarized as follows:
\begin{itemize}
    \item EP240617a is likely an untriggered GRB, in which early weak signals, close to the count rate corresponding to SNR = 2, can be identified through the average orbital background subtraction method. These signals are consistent with the light curve observed by EP-WXT.
    \item 
    {Due to observational limitations, the distance of this source cannot be determined, and the presence of an associated supernova or kilonova, which would serve as the smoking gun of its progenitor, has not been firmly established.}
    Given its X-ray to $\gamma$-ray fluence ratio, energy peak, and spectral index, It may be classified as an event intermediate between XRR and XRF.
\end{itemize}

Due to the fact that, in the early stage of EP’s scientific operation, the follow-up observation procedure of FXT had not yet been fully established, the follow-up data for EP240617a are rather limited. Nevertheless, events detected by EP-WXT with fluxes below the typical triggering threshold of $\gamma$-ray detectors are crucial for probing GRB-related physical processes. 
\rj{Events like EP240617a may prompt a re-evaluation of the trigger time of burst and the launching time of jet. In the Fermi observation of EP240617a, a weak signal could be identified approximately 140 seconds before the start of the $T_{90}$ interval defined by GBM. Obviously, without EP-WXT detection or orbit-averaged background analysis, this early emission might have been missed. In earlier observations of GRBs, precursor emission was already recognized as a common phenomenon \citep{1995ApJ...452..145K}. This burst suggests that the actual GRB start time may precede the conventionally defined trigger time. With access to lower-energy observational windows, such as those provided by WXT or even SVOM Ground-based Wide Angle Camera array (GWAC) \citep{2021PASP..133f5001H, 2023NatAs...7..724X}, the determination of the start time can be more accurate.}


\rj{Besides, EP240617a exhibits a continuing smooth emission phase detected by EP-WXT. The unusual behavior contrasts markedly with 
GRB 240315C \citep{2025NatAs...9..564L}, whose observed characteristics align more closely with either the ``tip-of-iceberg'' phenomenon \citep{2014MNRAS.442.1922L} or spectral evolution patterns. 
One of the possible explanations for the smooth plateau observed in EP240617a is the contribution from early external shocks, for which evidence has been observed in the BOAT GRB 221009A \citep{2023ApJ...956L..21Z} as well as in several other GRBs, such as GRB 120729A \citep{2018ApJ...859..163H} and GRB 200829A \citep{2023ApJ...944...21L}.
Another possible explanation is that the event may have originated from a neutron star-white dwarf (NS-WD) merger that launched a relativistic jet.}

\rj{Self-consistent magnetohydrodynamic simulations demonstrated by \cite{2024A&A...681A..41M} show a quite similar temporal evolution profile between their jet luminosity and the light curve of EP240617a.
The simulation results indicate the existence of a sustained quasi-steady accretion phase lasting on the order of hundreds of seconds during the early merger stage. 
Subsequently, rapid angular momentum loss, dominated by spiral density waves transporting angular momentum, ultimately drives the rapid accretion of WD material. 
This accretion process is accompanied by rapid amplification of the magnetic field of NS, eventually powering jet production.
Although \cite{2024A&A...681A..41M} acknowledge the inability to fully simulate the detailed behavior of the accretion flow near the NS, they however note the successful jets are produced prior to the rapid accretion phase. 
The authors characterize jet with luminosities of $\sim 10^{47}$ erg s$^{-1}$, duration on $\sim 10^3$ seconds, and a predicted maximum Lorentz factor of $\sim 10$. 
Assuming EP240617a indeed has a relatively low redshift, we find that some relations could be satisfied under these conditions. For example, the location in the the $E_{\rm p,z}$–$E_{\gamma,\mathrm{iso}}$ relation would suggest a local universe origin of EP240617a in order to be consistent with the region of a Type I GRBs.}

\rj{In conclusion, we reaffirm that the WD-NS merger scenario presents a viable framework for explaining some nearby low-luminosity GRBs. This model demonstrates particular relevance for events exhibiting: (1) prolonged precursor activity, (2) delayed rapid accretion signatures, and (3) moderated relativistic outflow properties - characteristics notably observed in EP240617a. The magnetic field amplification mechanism during the extended accretion phase appears crucial for bridging the temporal gap between initial merger dynamics and eventual relativistic jet launching.}

Despite LIGO/Virgo/KAGRA \citep{2018LRR....21....3A} being operational during the O4 run at the time of EP240617a, detecting NS-WD mergers necessitates next-generation gravitational wave detectors with enhanced sensitivity at low frequencies, such as TianQin \citep{2016CQGra..33c5010L}, DECIGO and BBO \citep{2011PhRvD..83d4011Y}, see also \citet{2023ApJ...954L..17Y} for more discussion on detector sensitivity. Coordinated space- and ground-based multi-messenger, multi-wavelength observations in the future will be essential for unveiling the nature of such events.

With the ongoing advancement of multi-messenger and multi-wavelength follow-up capabilities, {\tt RapidGBM} can be employed when wide-field instruments such as EP-WXT trigger on an event to promptly analyze the corresponding GBM data and assess whether the event is a GRB. At the same time, it can utilize empirical relations among prompt emission parameters to identify special GRB subclasses, such as long-short bursts GRB~211211A \citep{2022Natur.612..232Y}, GRB~230307A \citep{2024Natur.626..737L,2023ApJ...953L...8W}, and magnetar giant flares GRB~200415A \citep{2020ApJ...899..106Y}, GRB~231115A \citep{2024ApJ...969..127W}, providing valuable references for follow-up observations within the transient community. 

\section*{Acknowledgments}
Y.W. is supported by the Jiangsu Funding Program for Excellent Postdoctoral Talent (grant No.~2024ZB110), the Postdoctoral Fellowship Program (grant No.~GZC20241916) and the General Fund (grant No.~2024M763531) of the China Postdoctoral Science Foundation.
D.M.W. is supported by the
Strategic Priority Research Program of the Chinese Academy
of Sciences (grant No. XDB0550400), the National Key
Research and Development Program of China (Nos.
2024YFA1611704), and the National
Natural Science Foundation of China (NSFC; Nos. 12473049).
J.R. is supported by the Postdoctoral Innovation Talents Support Program (No.~BX20250160), the General Fund (Grant No.~2024M763530) of the China Postdoctoral Science Foundation, and the Jiangsu Funding Program for Excellent Postdoctoral Talent (grant No.~2025ZB272).
H.Z. is supported by the Postdoctoral Innovation Talents Support Program ( No.~BX20250159).
Z.P.J. is supported by the NSFC under No. 12225305.
This work is based on data obtained with the Einstein Probe, a space mission supported by the Strategic Priority Program on Space Science of the Chinese Academy of Sciences, in collaboration with ESA, MPE and CNES (Grant No. XDA15310000, No. XDA15052100). We also acknowledge the use of the Fermi archive's public data.
Y.W. thanks W. J. Xie, C. Wu, and L. P. Xin of the SVOM team for testing and helpful suggestions on the tool, and H. Q. Chen for support in EP-WXT data analysis.

\vspace{5mm}

\bibliography{sample631}{}

\begin{thebibliography}{}
\expandafter\ifx\csname natexlab\endcsname\relax\def\natexlab#1{#1}\fi
\providecommand{\url}[1]{\href{#1}{#1}}
\providecommand{\dodoi}[1]{doi:~\href{http://doi.org/#1}{\nolinkurl{#1}}}
\providecommand{\doeprint}[1]{\href{http://ascl.net/#1}{\nolinkurl{http://ascl.net/#1}}}
\providecommand{\doarXiv}[1]{\href{https://arxiv.org/abs/#1}{\nolinkurl{https://arxiv.org/abs/#1}}}

\bibitem[{{Abbott} {et~al.}(2018){Abbott}, {Abbott}, {Abbott}, {Abernathy},
  {Acernese}, {Ackley}, {Adams}, {Adams}, {Addesso}, {Adhikari}, {Adya},
  {Affeldt}, {Agathos}, {Agatsuma}, {Aggarwal}, {Aguiar}, {Aiello}, {Ain},
  {Ajith}, {Akutsu}, {Allen}, {Allocca}, {Altin}, {Ananyeva}, {Anderson},
  {Anderson}, {Ando}, {Appert}, {Arai}, {Araya}, {Araya}, {Areeda}, {Arnaud},
  {Arun}, {Asada}, {Ascenzi}, {Ashton}, {Aso}, {Ast}, {Aston}, {Astone},
  {Atsuta}, {Aufmuth}, {Aulbert}, {Avila-Alvarez}, {Awai}, {Babak}, {Bacon},
  {Bader}, {Baiotti}, {Baker}, {Baldaccini}, {Ballardin}, {Ballmer},
  {Barayoga}, {Barclay}, {Barish}, {Barker}, {Barone}, {Barr}, {Barsotti},
  {Barsuglia}, {Barta}, {Bartlett}, {Barton}, {Bartos}, {Bassiri}, {Basti},
  {Batch}, {Baune}, {Bavigadda}, {Bazzan}, {B{\'e}csy}, {Beer}, {Bejger},
  {Belahcene}, {Belgin}, {Bell}, {Berger}, {Bergmann}, {Berry}, {Bersanetti},
  {Bertolini}, {Betzwieser}, {Bhagwat}, {Bhandare}, {Bilenko}, {Billingsley},
  {Billman}, {Birch}, {Birney}, {Birnholtz}, {Biscans}, {Bisht}, {Bitossi},
  {Biwer}, {Bizouard}, {Blackburn}, {Blackman}, {Blair}, {Blair}, {Blair},
  {Bloemen}, {Bock}, {Boer}, {Bogaert}, {Bohe}, {Bondu}, {Bonnand}, {Boom},
  {Bork}, {Boschi}, {Bose}, {Bouffanais}, {Bozzi}, {Bradaschia}, {Brady},
  {Braginsky}, {Branchesi}, {Brau}, {Briant}, {Brillet}, {Brinkmann},
  {Brisson}, {Brockill}, {Broida}, {Brooks}, {Brown}, {Brown}, {Brown},
  {Brunett}, {Buchanan}, {Buikema}, {Bulik}, {Bulten}, {Buonanno}, {Buskulic},
  {Buy}, {Byer}, {Cabero}, {Cadonati}, {Cagnoli}, {Cahillane}, {Calder{\'o}n
  Bustillo}, {Callister}, {Calloni}, {Camp}, {Cannon}, {Cao}, {Cao}, {Capano},
  {Capocasa}, {Carbognani}, {Caride}, {Casanueva Diaz}, {Casentini}, {Caudill},
  {Cavagli{\`a}}, {Cavalier}, {Cavalieri}, {Cella}, {Cepeda}, {Cerboni
  Baiardi}, {Cerretani}, {Cesarini}, {Chamberlin}, {Chan}, {Chao}, {Charlton},
  {Chassande-Mottin}, {Cheeseboro}, {Chen}, {Chen}, {Cheng}, {Chincarini},
  {Chiummo}, {Chmiel}, {Cho}, {Cho}, {Chow}, {Christensen}, {Chu}, {Chua},
  {Chua}, {Chung}, {Ciani}, {Clara}, {Clark}, {Cleva}, {Cocchieri}, {Coccia},
  {Cohadon}, {Colla}, {Collette}, {Cominsky}, {Constancio}, {Conti}, {Cooper},
  {Corbitt}, \& {Cornish}}]{2018LRR....21....3A}
{Abbott}, B.~P., {Abbott}, R., {Abbott}, T.~D., {et~al.} 2018, Living Reviews
  in Relativity, 21, 3, \dodoi{10.1007/s41114-018-0012-9}

\bibitem[{{Ajello} {et~al.}(2014){Ajello}, {Albert}, {Allafort}, {Baldini},
  {Barbiellini}, {Bastieri}, {Bellazzini}, {Bissaldi}, {Bonamente}, {Brandt},
  {Bregeon}, {Brigida}, {Bruel}, {Buehler}, {Buson}, {Caliandro}, {Cameron},
  {Caraveo}, {Cecchi}, {Charles}, {Chekhtman}, {Chiang}, {Chiaro}, {Ciprini},
  {Claus}, {Cohen-Tanugi}, {Cominsky}, {Conrad}, {Cutini}, {D'Ammando}, {de
  Palma}, {Dermer}, {Desiante}, {Digel}, {Silva}, {Drell}, {Drlica-Wagner},
  {Favuzzi}, {Focke}, {Franckowiak}, {Fukazawa}, {Fusco}, {Gargano},
  {Gasparrini}, {Germani}, {Giglietto}, {Giommi}, {Giordano}, {Giroletti},
  {Glanzman}, {Godfrey}, {Grenier}, {Grove}, {Guiriec}, {Hadasch}, {Hayashida},
  {Hays}, {Horan}, {Hou}, {Hughes}, {Inoue}, {Jackson}, {Jogler},
  {J{\'o}hannesson}, {Johnson}, {Johnson}, {Kamae}, {Kn{\"o}dlseder},
  {Kocevski}, {Kuss}, {Lande}, {Larsson}, {Latronico}, {Longo}, {Loparco},
  {Lott}, {Lovellette}, {Lubrano}, {Mayer}, {Mazziotta}, {McEnery},
  {Michelson}, {Mizuno}, {Moiseev}, {Monte}, {Monzani}, {Morselli},
  {Moskalenko}, {Murgia}, {Murphy}, {Nakamori}, {Nemmen}, {Nuss}, {Ohno},
  {Ohsugi}, {Omodei}, {Orienti}, {Orlando}, {Ormes}, {Paneque}, {Panetta},
  {Perkins}, {Pesce-Rollins}, {Petrosian}, {Piron}, {Pivato}, {Porter},
  {Rain{\`o}}, {Rando}, {Razzano}, {Reimer}, {Reimer}, {Roth}, {Schulz},
  {Sgr{\`o}}, {Siskind}, {Spandre}, {Spinelli}, {Takahashi}, {Thayer},
  {Thayer}, {Thompson}, {Tibaldo}, {Tinivella}, {Tosti}, {Troja}, {Usher},
  {Vandenbroucke}, {Vasileiou}, {Vianello}, {Vitale}, {Werner}, {Winer},
  {Wood}, {Wood}, \& {Yang}}]{2014ApJ...789...20A}
{Ajello}, M., {Albert}, A., {Allafort}, A., {et~al.} 2014, \apj, 789, 20,
  \dodoi{10.1088/0004-637X/789/1/20}

\bibitem[{{Amati} {et~al.}(2002){Amati}, {Frontera}, {Tavani}, {in't Zand},
  {Antonelli}, {Costa}, {Feroci}, {Guidorzi}, {Heise}, {Masetti}, {Montanari},
  {Nicastro}, {Palazzi}, {Pian}, {Piro}, \& {Soffitta}}]{amati2002intrinsic}
{Amati}, L., {Frontera}, F., {Tavani}, M., {et~al.} 2002, \aap, 390, 81,
  \dodoi{10.1051/0004-6361:20020722}

\bibitem[{{Atteia} {et~al.}(2004){Atteia}, {Barraud}, {Lestrade}, {Olive},
  {Dezalay}, {Boer}, {Hurley}, {Jernigan}, {Ricker}, {Vanderspek}, {Crew},
  {Doty}, {Monnelly}, {Villasenor}, {Butler}, {Levine}, {Kawai}, {Yoshida},
  {Sakamoto}, {Tamagawa}, {Torii}, {Matsuoka}, {Lamb}, {Donaghy}, {Graziani},
  {Pizzichini}, {Fenimore}, {Galassi}, \& {Tavenner}}]{2004ASPC..312...12A}
{Atteia}, J.~L., {Barraud}, C., {Lestrade}, J.~P., {et~al.} 2004, in
  Astronomical Society of the Pacific Conference Series, Vol. 312, Gamma-Ray
  Bursts in the Afterglow Era, ed. M.~{Feroci}, F.~{Frontera}, N.~{Masetti}, \&
  L.~{Piro}, 12

\bibitem[{{Barraud} {et~al.}(2005){Barraud}, {Daigne}, {Mochkovitch}, \&
  {Atteia}}]{2005A&A...440..809B}
{Barraud}, C., {Daigne}, F., {Mochkovitch}, R., \& {Atteia}, J.~L. 2005, \aap,
  440, 809, \dodoi{10.1051/0004-6361:20041572}

\bibitem[{{Barraud} {et~al.}(2003){Barraud}, {Olive}, {Lestrade}, {Atteia},
  {Hurley}, {Ricker}, {Lamb}, {Kawai}, {Boer}, {Dezalay}, {Pizzichini},
  {Vanderspek}, {Crew}, {Doty}, {Monnelly}, {Villasenor}, {Butler}, {Levine},
  {Yoshida}, {Shirasaki}, {Sakamoto}, {Tamagawa}, {Torii}, {Matsuoka},
  {Fenimore}, {Galassi}, {Tavenner}, {Donaghy}, {Graziani}, \&
  {Jernigan}}]{2003A&A...400.1021B}
{Barraud}, C., {Olive}, J.~F., {Lestrade}, J.~P., {et~al.} 2003, \aap, 400,
  1021, \dodoi{10.1051/0004-6361:20030074}

\bibitem[{{Berger}(2014)}]{2014ARA&A..52...43B}
{Berger}, E. 2014, \araa, 52, 43, \dodoi{10.1146/annurev-astro-081913-035926}

\bibitem[{{Bloom} {et~al.}(2001){Bloom}, {Frail}, \&
  {Sari}}]{2001AJ....121.2879B}
{Bloom}, J.~S., {Frail}, D.~A., \& {Sari}, R. 2001, \aj, 121, 2879,
  \dodoi{10.1086/321093}

\bibitem[{{Cheng} {et~al.}(2025){Cheng}, {Zhang}, {Ling}, {Sun}, {Sun}, {Liu},
  {Dai}, {Jia}, {Pan}, {Wang}, {Zhao}, {Chen}, {Cheng}, {Fu}, {Han}, {Li},
  {Li}, {Ma}, {Xue}, {Yan}, {Zhang}, {Wang}, {Yang}, {Zhao}, {Li}, {Jin}, \&
  {Yuan}}]{Cheng2025}
{Cheng}, H., {Zhang}, C., {Ling}, Z., {et~al.} 2025, arXiv e-prints,
  arXiv:2505.18939.
\newblock \doarXiv{2505.18939}

\bibitem[{{Eichler} {et~al.}(1996){Eichler}, {Livio}, {Piran}, \&
  {Schramm}}]{1996bboe.book..682E}
{Eichler}, D., {Livio}, M., {Piran}, T., \& {Schramm}, D.~N. 1996, in The Big
  Bang and Other Explosions in Nuclear and Particle Astrophysics. Edited by
  SCHRAMM DAVID N. Published by World Scientific Publishing Co. Pte. Ltd,
  682--684, \dodoi{10.1142/9789812831538_0076}

\bibitem[{{Fitzpatrick} {et~al.}(2011){Fitzpatrick}, {Connaughton}, {McBreen},
  \& {Tierney}}]{2011arXiv1111.3779F}
{Fitzpatrick}, G., {Connaughton}, V., {McBreen}, S., \& {Tierney}, D. 2011,
  arXiv e-prints, arXiv:1111.3779, \dodoi{10.48550/arXiv.1111.3779}

\bibitem[{{Fitzpatrick} {et~al.}(2012){Fitzpatrick}, {McBreen}, {Connaughton},
  \& {Briggs}}]{2012SPIE.8443E..3BF}
{Fitzpatrick}, G., {McBreen}, S., {Connaughton}, V., \& {Briggs}, M. 2012, in
  Society of Photo-Optical Instrumentation Engineers (SPIE) Conference Series,
  Vol. 8443, Space Telescopes and Instrumentation 2012: Ultraviolet to Gamma
  Ray, ed. T.~{Takahashi}, S.~S. {Murray}, \& J.-W.~A. {den Herder}, 84433B,
  \dodoi{10.1117/12.928036}

\bibitem[{{Fong} \& {Berger}(2013)}]{2013ApJ...776...18F}
{Fong}, W., \& {Berger}, E. 2013, \apj, 776, 18,
  \dodoi{10.1088/0004-637X/776/1/18}

\bibitem[{{Fong} {et~al.}(2010){Fong}, {Berger}, \&
  {Fox}}]{2010ApJ...708....9F}
{Fong}, W., {Berger}, E., \& {Fox}, D.~B. 2010, \apj, 708, 9,
  \dodoi{10.1088/0004-637X/708/1/9}

\bibitem[{{Fruchter} {et~al.}(2006){Fruchter}, {Levan}, {Strolger},
  {Vreeswijk}, {Thorsett}, {Bersier}, {Burud}, {Castro Cer{\'o}n},
  {Castro-Tirado}, {Conselice}, {Dahlen}, {Ferguson}, {Fynbo}, {Garnavich},
  {Gibbons}, {Gorosabel}, {Gull}, {Hjorth}, {Holland}, {Kouveliotou}, {Levay},
  {Livio}, {Metzger}, {Nugent}, {Petro}, {Pian}, {Rhoads}, {Riess}, {Sahu},
  {Smette}, {Tanvir}, {Wijers}, \& {Woosley}}]{2006Natur.441..463F}
{Fruchter}, A.~S., {Levan}, A.~J., {Strolger}, L., {et~al.} 2006, \nat, 441,
  463, \dodoi{10.1038/nature04787}

\bibitem[{{Gehrels} {et~al.}(2005){Gehrels}, {Sarazin}, {O'Brien}, {Zhang},
  {Barbier}, {Barthelmy}, {Blustin}, {Burrows}, {Cannizzo}, {Cummings}, {Goad},
  {Holland}, {Hurkett}, {Kennea}, {Levan}, {Markwardt}, {Mason}, {Meszaros},
  {Page}, {Palmer}, {Rol}, {Sakamoto}, {Willingale}, {Angelini}, {Beardmore},
  {Boyd}, {Breeveld}, {Campana}, {Chester}, {Chincarini}, {Cominsky},
  {Cusumano}, {de Pasquale}, {Fenimore}, {Giommi}, {Gronwall}, {Grupe}, {Hill},
  {Hinshaw}, {Hjorth}, {Hullinger}, {Hurley}, {Klose}, {Kobayashi},
  {Kouveliotou}, {Krimm}, {Mangano}, {Marshall}, {McGowan}, {Moretti},
  {Mushotzky}, {Nakazawa}, {Norris}, {Nousek}, {Osborne}, {Page}, {Parsons},
  {Patel}, {Perri}, {Poole}, {Romano}, {Roming}, {Rosen}, {Sato}, {Schady},
  {Smale}, {Sollerman}, {Starling}, {Still}, {Suzuki}, {Tagliaferri},
  {Takahashi}, {Tashiro}, {Tueller}, {Wells}, {White}, \&
  {Wijers}}]{2005Natur.437..851G}
{Gehrels}, N., {Sarazin}, C.~L., {O'Brien}, P.~T., {et~al.} 2005, \nat, 437,
  851, \dodoi{10.1038/nature04142}

\bibitem[{Goldstein {et~al.}(2021)Goldstein, Cleveland, \&
  Kocevski}]{GbmDataTools}
Goldstein, A., Cleveland, W.~H., \& Kocevski, D. 2021, Fermi GBM Data Tools:
  v1.1.0.
\newblock \url{https://fermi.gsfc.nasa.gov/ssc/data/analysis/gbm}

\bibitem[{{Han} {et~al.}(2021){Han}, {Xiao}, {Zhang}, {Turpin}, {Xin}, {Wu},
  {Cai}, {Dong}, {Huang}, {Kang}, {Leroy}, {Li}, {Li}, {Lu}, {Qiu}, {Stahl},
  {Wang}, {Wang}, {Xu}, {Yang}, {Zhao}, {Zhang}, {Zheng}, {Zheng}, \&
  {Wei}}]{2021PASP..133f5001H}
{Han}, X., {Xiao}, Y., {Zhang}, P., {et~al.} 2021, \pasp, 133, 065001,
  \dodoi{10.1088/1538-3873/abfb4e}

\bibitem[{{Heise} {et~al.}(2001){Heise}, {Zand}, {Kippen}, \&
  {Woods}}]{2001grba.conf...16H}
{Heise}, J., {Zand}, J.~I., {Kippen}, R.~M., \& {Woods}, P.~M. 2001, in
  Gamma-ray Bursts in the Afterglow Era, ed. E.~{Costa}, F.~{Frontera}, \&
  J.~{Hjorth}, 16, \dodoi{10.1007/10853853_4}

\bibitem[{{Huang} {et~al.}(2018){Huang}, {Wang}, {Zheng}, {Liang}, {Lin},
  {Zhong}, {Zhang}, {Huang}, {Filippenko}, \& {Zhang}}]{2018ApJ...859..163H}
{Huang}, L.-Y., {Wang}, X.-G., {Zheng}, W., {et~al.} 2018, \apj, 859, 163,
  \dodoi{10.3847/1538-4357/aaba6e}

\bibitem[{Jeffreys(1998)}]{jeffreys1998theory}
Jeffreys, H. 1998, The theory of probability (OuP Oxford)

\bibitem[{{Jiang} {et~al.}(2025){Jiang}, {Xu}, {van Hoof}, {Lei}, {Liu},
  {Zhou}, {Chen}, {Fu}, {Yang}, {Liu}, {Zhu}, {Filippenko}, {Jonker},
  {Pozanenko}, {Gao}, {Wu}, {Zhang}, {Lamb}, {De Pasquale}, {Kobayashi},
  {Bauer}, {Sun}, {Pugliese}, {An}, {D'Elia}, {Fynbo}, {Zheng}, {Tirado},
  {Yin}, {Zou}, {Deller}, {Pankov}, {Volnova}, {Moskvitin}, {Spiridonova},
  {Oparin}, {Rumyantsev}, {Burkhonov}, {Egamberdiyev}, {Kim}, {Krugov},
  {Tatarnikov}, {Inasaridze}, {Levan}, {Bj{\o}rn Malesani}, {Ravasio},
  {Quirola-V{\'a}squez}, {van Dalen}, {S{\'a}nchez-Sierras}, {Mata
  S{\'a}nchez}, {Littlefair}, {Chac{\'o}n}, {Torres}, {Chrimes}, {Sarin},
  {Martin-Carrillo}, {Dhillon}, {Yang}, {Brink}, {Davies}, {Yang}, {Aryan},
  {Chen}, {Kong}, {Li}, {Li}, {Mao}, {P{\'e}rez-Garc{\'\i}a},
  {Fern{\'a}ndez-Garc{\'\i}a}, {Andrews}, {Farah}, {Fan}, {Padilla Gonzalez},
  {Howell}, {Hartmann}, {Hu}, {Jakobsson}, {Li}, {Ling}, {McCully}, {Newsome},
  {Schneider}, {Samaporn Tinyanont}, {Sun}, {Terreran}, {Tang}, {Wang}, {Xu},
  {Yuan}, {Zhang}, {Zhao}, \& {Zhang}}]{2025arXiv250304306J}
{Jiang}, S.-Q., {Xu}, D., {van Hoof}, A. P.~C., {et~al.} 2025, arXiv e-prints,
  arXiv:2503.04306, \dodoi{10.48550/arXiv.2503.04306}

\bibitem[{{Kaneko} {et~al.}(2015){Kaneko}, {Bostanc{\i}},
  {G{\"o}{\u{g}}{\"u}{\c{s}}}, \& {Lin}}]{2015MNRAS.452..824K}
{Kaneko}, Y., {Bostanc{\i}}, Z.~F., {G{\"o}{\u{g}}{\"u}{\c{s}}}, E., \& {Lin},
  L. 2015, \mnras, 452, 824, \dodoi{10.1093/mnras/stv1286}

\bibitem[{{Kippen} {et~al.}(2003){Kippen}, {Woods}, {Heise}, {in't Zand},
  {Briggs}, \& {Preece}}]{2003AIPC..662..244K}
{Kippen}, R.~M., {Woods}, P.~M., {Heise}, J., {et~al.} 2003, in American
  Institute of Physics Conference Series, Vol. 662, Gamma-Ray Burst and
  Afterglow Astronomy 2001: A Workshop Celebrating the First Year of the HETE
  Mission, ed. G.~R. {Ricker} \& R.~K. {Vanderspek} (AIP), 244--247,
  \dodoi{10.1063/1.1579349}

\bibitem[{{Koshut} {et~al.}(1995){Koshut}, {Kouveliotou}, {Paciesas}, {van
  Paradijs}, {Pendleton}, {Briggs}, {Fishman}, \&
  {Meegan}}]{1995ApJ...452..145K}
{Koshut}, T.~M., {Kouveliotou}, C., {Paciesas}, W.~S., {et~al.} 1995, \apj,
  452, 145, \dodoi{10.1086/176286}

\bibitem[{{Koshut} {et~al.}(1996){Koshut}, {Paciesas}, {Kouveliotou}, {van
  Paradijs}, {Pendleton}, {Fishman}, \& {Meegan}}]{koshut1996systematic}
{Koshut}, T.~M., {Paciesas}, W.~S., {Kouveliotou}, C., {et~al.} 1996, \apj,
  463, 570, \dodoi{10.1086/177272}

\bibitem[{{Leibler} \& {Berger}(2010)}]{2010ApJ...725.1202L}
{Leibler}, C.~N., \& {Berger}, E. 2010, \apj, 725, 1202,
  \dodoi{10.1088/0004-637X/725/1/1202}

\bibitem[{{Lesage} {et~al.}(2023){Lesage}, {Veres}, {Briggs}, {Goldstein},
  {Kocevski}, {Burns}, {Wilson-Hodge}, {Bhat}, {Huppenkothen}, {Fryer},
  {Hamburg}, {Racusin}, {Bissaldi}, {Cleveland}, {Dalessi}, {Fletcher},
  {Giles}, {Hristov}, {Hui}, {Mailyan}, {Malacaria}, {Poolakkil}, {Roberts},
  {von Kienlin}, {Wood}, {Ajello}, {Arimoto}, {Baldini}, {Ballet}, {Baring},
  {Bastieri}, {Gonzalez}, {Bellazzini}, {Bissaldi}, {Blandford}, {Bonino},
  {Bruel}, {Buson}, {Cameron}, {Caputo}, {Caraveo}, {Cavazzuti}, {Chiaro},
  {Cibrario}, {Ciprini}, {Orestano}, {Crnogorcevic}, {Cuoco}, {Cutini},
  {D'Ammando}, {De Gaetano}, {Di Lalla}, {Di Venere}, {Dom{\'\i}nguez},
  {Fegan}, {Ferrara}, {Fleischhack}, {Fukazawa}, {Funk}, {Fusco}, {Galanti},
  {Gammaldi}, {Gargano}, {Gasbarra}, {Gasparrini}, {Germani}, {Giacchino},
  {Giglietto}, {Gill}, {Giroletti}, {Granot}, {Green}, {Grenier}, {Guiriec},
  {Gustafsson}, {Hays}, {Hewitt}, {Horan}, {Hou}, {Kuss}, {Latronico},
  {Laviron}, {Lemoine-Goumard}, {Li}, {Liodakis}, {Longo}, {Loparco},
  {Lorusso}, {Lovellette}, {Lubrano}, {Maldera}, {Manfreda},
  {Mart{\'\i}-Devesa}, {Mazziotta}, {McEnery}, {Mereu}, {Meyer}, {Michelson},
  {Mizuno}, {Monzani}, {Morselli}, {Moskalenko}, {Negro}, {Nuss}, {Omodei},
  {Orlando}, {Ormes}, {Paneque}, {Panzarini}, {Persic}, {Pesce-Rollins},
  {Pillera}, {Piron}, {Poon}, {Porter}, {Principe}, {Rain{\`o}}, {Rando},
  {Rani}, {Razzano}, {Razzaque}, {Reimer}, {Reimer}, {Ryde},
  {S{\'a}nchez-Conde}, {Parkinson}, {Scotton}, {Serini}, {Sgr{\`o}}, {Sharma},
  {Siskind}, {Spandre}, {Spinelli}, {Tajima}, {Torres}, {Valverde}, {Venters},
  {Wadiasingh}, {Wood}, \& {Zaharijas}}]{2023ApJ...952L..42L}
{Lesage}, S., {Veres}, P., {Briggs}, M.~S., {et~al.} 2023, \apjl, 952, L42,
  \dodoi{10.3847/2041-8213/ace5b4}

\bibitem[{{Levan} {et~al.}(2024{\natexlab{a}}){Levan}, {Jonker}, {Saccardi},
  {Bj{\o}rn Malesani}, {Tanvir}, {Izzo}, {Heintz}, {Mata S{\'a}nchez},
  {Quirola-V{\'a}squez}, {Torres}, {Vergani}, {Schulze}, {Rossi}, {D'Avanzo},
  {Gompertz}, {Martin-Carrillo}, {de Ugarte Postigo}, {Schneider}, {Yuan},
  {Ling}, {Zhang}, {Mao}, {Liu}, {Sun}, {Xu}, {Zhu}, {Ag{\"u}{\'\i}
  Fern{\'a}ndez}, {Amati}, {Bauer}, {Campana}, {Carotenuto}, {Chrimes}, {van
  Dalen}, {D'Elia}, {Della Valle}, {De Pasquale}, {Dhillon}, {Galbany},
  {Gaspari}, {Gianfagna}, {Gomboc}, {Habeeb}, {van Hoof}, {Hu}, {Jakobsson},
  {Julakanti}, {Korth}, {Kouveliotou}, {Laskar}, {Littlefair}, {Maiorano},
  {Mao}, {Melandri}, {Miller}, {Mukherjee}, {Oates}, {O'Brien}, {Palmerio},
  {Parviainen}, {Pieterse}, {Piranomonte}, {Piro}, {Pugliese}, {Ravasio},
  {Rayson}, {Salvaterra}, {S{\'a}nchez-Ram{\'\i}rez}, {Sarin}, {Shilling},
  {Starling}, {Tagliaferri}, {Linesh Thakur}, {Th{\"o}ne}, {Wiersema},
  {Worssam}, \& {Zafar}}]{2024arXiv240416350L}
{Levan}, A.~J., {Jonker}, P.~G., {Saccardi}, A., {et~al.} 2024{\natexlab{a}},
  arXiv e-prints, arXiv:2404.16350, \dodoi{10.48550/arXiv.2404.16350}

\bibitem[{{Levan} {et~al.}(2024{\natexlab{b}}){Levan}, {Gompertz}, {Salafia},
  {Bulla}, {Burns}, {Hotokezaka}, {Izzo}, {Lamb}, {Malesani}, {Oates},
  {Ravasio}, {Rouco Escorial}, {Schneider}, {Sarin}, {Schulze}, {Tanvir},
  {Ackley}, {Anderson}, {Brammer}, {Christensen}, {Dhillon}, {Evans},
  {Fausnaugh}, {Fong}, {Fruchter}, {Fryer}, {Fynbo}, {Gaspari}, {Heintz},
  {Hjorth}, {Kennea}, {Kennedy}, {Laskar}, {Leloudas}, {Mandel},
  {Martin-Carrillo}, {Metzger}, {Nicholl}, {Nugent}, {Palmerio}, {Pugliese},
  {Rastinejad}, {Rhodes}, {Rossi}, {Saccardi}, {Smartt}, {Stevance},
  {Tohuvavohu}, {van der Horst}, {Vergani}, {Watson}, {Barclay},
  {Bhirombhakdi}, {Breedt}, {Breeveld}, {Brown}, {Campana}, {Chrimes},
  {D'Avanzo}, {D'Elia}, {De Pasquale}, {Dyer}, {Galloway}, {Garbutt}, {Green},
  {Hartmann}, {Jakobsson}, {Kerry}, {Kouveliotou}, {Langeroodi}, {Le Floc'h},
  {Leung}, {Littlefair}, {Munday}, {O'Brien}, {Parsons}, {Pelisoli}, {Sahman},
  {Salvaterra}, {Sbarufatti}, {Steeghs}, {Tagliaferri}, {Th{\"o}ne}, {de Ugarte
  Postigo}, \& {Kann}}]{2024Natur.626..737L}
{Levan}, A.~J., {Gompertz}, B.~P., {Salafia}, O.~S., {et~al.}
  2024{\natexlab{b}}, \nat, 626, 737, \dodoi{10.1038/s41586-023-06759-1}

\bibitem[{{Li} {et~al.}(2023){Li}, {Lin}, {Lu}, {Jiang}, {Liang}, {Chen}, {Li},
  {Wang}, \& {Liang}}]{2023ApJ...944...21L}
{Li}, J., {Lin}, D.-B., {Lu}, R.-J., {et~al.} 2023, \apj, 944, 21,
  \dodoi{10.3847/1538-4357/acaf68}

\bibitem[{{Li} \& {Ma}(1983)}]{1983ApJ...272..317L}
{Li}, T.~P., \& {Ma}, Y.~Q. 1983, \apj, 272, 317, \dodoi{10.1086/161295}

\bibitem[{{Liu} {et~al.}(2025){Liu}, {Sun}, {Xu}, {Svinkin}, {Delaunay},
  {Tanvir}, {Gao}, {Zhang}, {Chen}, {Wu}, {Zhang}, {Yuan}, {An}, {Bruni},
  {Frederiks}, {Ghirlanda}, {Hu}, {Li}, {Li}, {Li}, {Malesani}, {Piro},
  {Raman}, {Ricci}, {Troja}, {Vergani}, {Wu}, {Yang}, {Zhang}, {Zhu}, {de
  Ugarte Postigo}, {Demin}, {Dobie}, {Fan}, {Fu}, {Fynbo}, {Geng}, {Gianfagna},
  {Hu}, {Huang}, {Jiang}, {Jonker}, {Julakanti}, {Kennea}, {Kokomov},
  {Kuulkers}, {Lei}, {Leung}, {Levan}, {Li}, {Li}, {Littlefair}, {Liu},
  {Lysenko}, {Ma}, {Martin-Carrillo}, {O'Brien}, {Parsotan},
  {Quirola-V{\'a}squez}, {Ridnaia}, {Ronchini}, {Rossi}, {Mata-S{\'a}nchez},
  {Schneider}, {Shen}, {Thakur}, {Tohuvavohu}, {Torres}, {Tsvetkova}, {Ulanov},
  {Wei}, {Xiao}, {Yin}, {Bai}, {Burwitz}, {Cai}, {Chen}, {Chen}, {Chen},
  {Chen}, {Chen}, {Chen}, {Cheng}, {Cordier}, {Cui}, {Cui}, {Dai}, {Dai},
  {Eder}, {Eyles-Ferris}, {Fan}, {Feldman}, {Feng}, {Feng}, {Friedrich}, {Gao},
  {Gonzalez}, {Guan}, {Han}, {Han}, {Hou}, {Hu}, {Hu}, {Huang}, {Huo},
  {Hutchinson}, {Ji}, {Jia}, {Jia}, {Jiang}, {Jin}, {Jin}, {Jin}, {Keereman},
  {Lerman}, {Li}, {Li}, {Li}, {Li}, {Li}, {Lian}, {Liang}, {Ling}, {Liu},
  {Liu}, {Liu}, {Liu}, {Liu}, {Lu}, {L{\"u}}, {Luo}, {Ma}, {Ma}, {Mao}, {Mao},
  {McHugh}, {Meidinger}, {Nandra}, {Osborne}, {Pan}, {Pan}, {Ravasio}, {Rau},
  {Rea}, {Rehman}, {Sanders}, {Santovincenzo}, {Song}, {Su}, {Sun}, {Sun},
  {Sun}, {Tan}, {Tang}, {Tao}, {Tong}, {Wang}, {Wang}, {Wang}, {Wang}, {Wang},
  {Wang}, {Wang}, {Wang}, {Wang}, {Wei}, {Willingale}, {Xiong}, {Xu}, {Xu},
  {Xu}, {Xu}, {Xu}, {Xue}, {Xue}, {Yan}, {Yang}, {Yang}, {Yang}, {Yang}, {Yu},
  {Zhang}, {Zhang}, {Zhang}, {Zhang}, {Zhang}, {Zhang}, {Zhang}, {Zhang},
  {Zhang}, {Zhao}, {Zhao}, {Zhao}, {Zhao}, {Zhou}, {Zhou}, {Zhu}, {Zhu}, \&
  {Zuo}}]{2025NatAs...9..564L}
{Liu}, Y., {Sun}, H., {Xu}, D., {et~al.} 2025, Nature Astronomy, 9, 564,
  \dodoi{10.1038/s41550-024-02449-8}

\bibitem[{{L{\"u}} {et~al.}(2014){L{\"u}}, {Zhang}, {Liang}, {Zhang}, \&
  {Sakamoto}}]{2014MNRAS.442.1922L}
{L{\"u}}, H.-J., {Zhang}, B., {Liang}, E.-W., {Zhang}, B.-B., \& {Sakamoto}, T.
  2014, \mnras, 442, 1922, \dodoi{10.1093/mnras/stu982}

\bibitem[{{Luo} {et~al.}(2016){Luo}, {Chen}, {Duan}, {Gong}, {Hu}, {Ji}, {Liu},
  {Mei}, {Milyukov}, {Sazhin}, {Shao}, {Toth}, {Tu}, {Wang}, {Wang}, {Yeh},
  {Zhan}, {Zhang}, {Zharov}, \& {Zhou}}]{2016CQGra..33c5010L}
{Luo}, J., {Chen}, L.-S., {Duan}, H.-Z., {et~al.} 2016, Classical and Quantum
  Gravity, 33, 035010, \dodoi{10.1088/0264-9381/33/3/035010}

\bibitem[{{Meegan} {et~al.}(2009){Meegan}, {Lichti}, {Bhat}, {Bissaldi},
  {Briggs}, {Connaughton}, {Diehl}, {Fishman}, {Greiner}, {Hoover}, {van der
  Horst}, {von Kienlin}, {Kippen}, {Kouveliotou}, {McBreen}, {Paciesas},
  {Preece}, {Steinle}, {Wallace}, {Wilson}, \&
  {Wilson-Hodge}}]{meegan2009fermi}
{Meegan}, C., {Lichti}, G., {Bhat}, P.~N., {et~al.} 2009, \apj, 702, 791,
  \dodoi{10.1088/0004-637X/702/1/791}

\bibitem[{{Minaev} \& {Pozanenko}(2020{\natexlab{a}})}]{minaev2020p}
{Minaev}, P.~Y., \& {Pozanenko}, A.~S. 2020{\natexlab{a}}, \mnras, 492, 1919,
  \dodoi{10.1093/mnras/stz3611}

\bibitem[{{Minaev} \& {Pozanenko}(2020{\natexlab{b}})}]{2020AstL...46..573M}
---. 2020{\natexlab{b}}, Astronomy Letters, 46, 573,
  \dodoi{10.1134/S1063773720090042}

\bibitem[{{Mor{\'a}n-Fraile} {et~al.}(2024){Mor{\'a}n-Fraile}, {R{\"o}pke},
  {Pakmor}, {Aloy}, {Ohlmann}, {Schneider}, {Leidi}, \&
  {Lioutas}}]{2024A&A...681A..41M}
{Mor{\'a}n-Fraile}, J., {R{\"o}pke}, F.~K., {Pakmor}, R., {et~al.} 2024, \aap,
  681, A41, \dodoi{10.1051/0004-6361/202347555}

\bibitem[{{Narayan} {et~al.}(1992){Narayan}, {Paczynski}, \&
  {Piran}}]{1992ApJ...395L..83N}
{Narayan}, R., {Paczynski}, B., \& {Piran}, T. 1992, \apjl, 395, L83,
  \dodoi{10.1086/186493}

\bibitem[{{Perez-Garcia} {et~al.}(2024){Perez-Garcia}, {Fernandez-Garcia},
  {Caballero-Garcia}, {Sanchez-Ramirez}, {Guziy}, {Wu}, {Castro-Tirado},
  {Meintjes}, {van Heerden}, {Martin-Carrillo}, {Hanlon}, {Gritsevich},
  {Xiong}, \& {Perez del Pulgar}}]{2024GCN.36693....1P}
{Perez-Garcia}, I., {Fernandez-Garcia}, E., {Caballero-Garcia}, M.~D., {et~al.}
  2024, GRB Coordinates Network, 36693, 1

\bibitem[{{Sakamoto} {et~al.}(2008){Sakamoto}, {Hullinger}, {Sato}, {Yamazaki},
  {Barbier}, {Barthelmy}, {Cummings}, {Fenimore}, {Gehrels}, {Krimm}, {Lamb},
  {Markwardt}, {Osborne}, {Palmer}, {Parsons}, {Stamatikos}, \&
  {Tueller}}]{2008ApJ...679..570S}
{Sakamoto}, T., {Hullinger}, D., {Sato}, G., {et~al.} 2008, \apj, 679, 570,
  \dodoi{10.1086/586884}

\bibitem[{{Santos} {et~al.}(2024){Santos}, {Bom}, {Kilpatrick},
  {Santana-Silva}, {Darc}, {Mendes de Oliveira}, \& {STEP
  Collaboration}}]{2024GCN.36707....1S}
{Santos}, A., {Bom}, C.~R., {Kilpatrick}, C.~D., {et~al.} 2024, GRB Coordinates
  Network, 36707, 1

\bibitem[{{Scargle} {et~al.}(2013){Scargle}, {Norris}, {Jackson}, \&
  {Chiang}}]{scargle2013studies}
{Scargle}, J.~D., {Norris}, J.~P., {Jackson}, B., \& {Chiang}, J. 2013, \apj,
  764, 167, \dodoi{10.1088/0004-637X/764/2/167}

\bibitem[{{Stratta} {et~al.}(2007){Stratta}, {Basa}, {Butler}, {Atteia},
  {Gendre}, {P{\'e}langeon}, {Malacrino}, {Mellier}, {Kann}, {Klose}, {Zeh},
  {Masetti}, {Palazzi}, {Gorosabel}, {Castro-Tirado}, {de Ugarte Postigo},
  {Jelinek}, {Cepa}, {Casta{\~n}eda}, {Mart{\'\i}nez-Delgado}, {Bo{\"e}r},
  {Braga}, {Crew}, {Donaghy}, {Dezalay}, {Doty}, {Fenimore}, {Galassi},
  {Graziani}, {Jernigan}, {Kawai}, {Lamb}, {Levine}, {Manchanda}, {Martel},
  {Matsuoka}, {Nakagawa}, {Olive}, {Pizzichini}, {Prigozhin}, {Ricker},
  {Sakamoto}, {Shirasaki}, {Sugita}, {Suzuki}, {Takagishi}, {Tamagawa},
  {Vanderspek}, {Villasenor}, {Woosley}, {Yamauchi}, \&
  {Yoshida}}]{2007A&A...461..485S}
{Stratta}, G., {Basa}, S., {Butler}, N., {et~al.} 2007, \aap, 461, 485,
  \dodoi{10.1051/0004-6361:20065831}

\bibitem[{{Sun} {et~al.}(2024){Sun}, {Chen}, {Zhou}, {Zhang}, {Li}, {Ling},
  {Liu}, {Zhang}, {Jin}, {Cheng}, {Cui}, {Fan}, {Hu}, {Hu}, {Huang}, {Liu},
  {Liu}, {Lv}, {Lian}, {Mao}, {Pan}, {Pan}, {Wang}, {Wang}, {Wu}, {Xu}, {Xu},
  {Yang}, {Yuan}, {Zhang}, {Zhang}, {Zhang}, {Zhang}, {Zhao}, {Chen}, {Jia},
  {Zhang}, {Kuulkers}, {Santovincenzo}, {O'Brien}, {Nandra}, {Rau}, {Cordier},
  \& {Einstein Probe Team}}]{2024GCN.36722....1S}
{Sun}, H., {Chen}, W., {Zhou}, H., {et~al.} 2024, GRB Coordinates Network,
  36722, 1

\bibitem[{{Thrane} \& {Talbot}(2019)}]{Thrane2019}
{Thrane}, E., \& {Talbot}, C. 2019, \pasa, 36, e010,
  \dodoi{10.1017/pasa.2019.2}

\bibitem[{{Wang}(2025)}]{RapidGBM}
{Wang}, Y. 2025, RapidGBM, 1.0,  Zenodo, \dodoi{10.5281/zenodo.16895816}

\bibitem[{{Wang} {et~al.}(2024){Wang}, {Wei}, {Zhou}, {Ren}, {Xia}, \&
  {Jin}}]{2024ApJ...969..127W}
{Wang}, Y., {Wei}, Y.-J., {Zhou}, H., {et~al.} 2024, \apj, 969, 127,
  \dodoi{10.3847/1538-4357/ad499f}

\bibitem[{{Wang} {et~al.}(2023){Wang}, {Xia}, {Zheng}, {Ren}, \&
  {Fan}}]{2023ApJ...953L...8W}
{Wang}, Y., {Xia}, Z.-Q., {Zheng}, T.-C., {Ren}, J., \& {Fan}, Y.-Z. 2023,
  \apjl, 953, L8, \dodoi{10.3847/2041-8213/ace7d4}

\bibitem[{{Willingale} {et~al.}(2013){Willingale}, {Starling}, {Beardmore},
  {Tanvir}, \& {O'Brien}}]{2013MNRAS.431..394W}
{Willingale}, R., {Starling}, R.~L.~C., {Beardmore}, A.~P., {Tanvir}, N.~R., \&
  {O'Brien}, P.~T. 2013, \mnras, 431, 394, \dodoi{10.1093/mnras/stt175}

\bibitem[{{Woosley}(1993)}]{1993ApJ...405..273W}
{Woosley}, S.~E. 1993, \apj, 405, 273, \dodoi{10.1086/172359}

\bibitem[{{Xin} {et~al.}(2023){Xin}, {Han}, {Li}, {Zhang}, {Wang}, {Turpin},
  {Yang}, {Qiu}, {Liang}, {Dai}, {Cai}, {Lu}, {Wang}, {Huang}, {Wang}, {Wu},
  {Gao}, {Ren}, {Zhang}, {Yang}, {Deng}, \& {Wei}}]{2023NatAs...7..724X}
{Xin}, L., {Han}, X., {Li}, H., {et~al.} 2023, Nature Astronomy, 7, 724,
  \dodoi{10.1038/s41550-023-01930-0}

\bibitem[{{Yagi} \& {Seto}(2011)}]{2011PhRvD..83d4011Y}
{Yagi}, K., \& {Seto}, N. 2011, \prd, 83, 044011,
  \dodoi{10.1103/PhysRevD.83.044011}

\bibitem[{{Yang} {et~al.}(2024){Yang}, {Yin}, {Zhang}, {Sun}, {Wu}, \&
  {Wu}}]{2024GCN.36692....1Y}
{Yang}, J., {Yin}, Y.-H.~I., {Zhang}, B., {et~al.} 2024, GRB Coordinates
  Network, 36692, 1

\bibitem[{{Yang} {et~al.}(2020){Yang}, {Chand}, {Zhang}, {Yang}, {Zou}, {Yang},
  {Zhao}, {Shao}, {Xiong}, {Luo}, {Li}, {Xiao}, {Li}, {Liu}, {Joshi}, {Sharma},
  {Chakraborty}, {Li}, \& {Zhang}}]{2020ApJ...899..106Y}
{Yang}, J., {Chand}, V., {Zhang}, B.-B., {et~al.} 2020, \apj, 899, 106,
  \dodoi{10.3847/1538-4357/aba745}

\bibitem[{{Yang} {et~al.}(2022){Yang}, {Ai}, {Zhang}, {Zhang}, {Liu}, {Wang},
  {Yang}, {Yin}, {Li}, \& {L{\"u}}}]{2022Natur.612..232Y}
{Yang}, J., {Ai}, S., {Zhang}, B.-B., {et~al.} 2022, \nat, 612, 232,
  \dodoi{10.1038/s41586-022-05403-8}

\bibitem[{{Yin} {et~al.}(2023){Yin}, {Zhang}, {Sun}, {Yang}, {Kang}, {Shao},
  {Yang}, \& {Zhang}}]{2023ApJ...954L..17Y}
{Yin}, Y.-H.~I., {Zhang}, B.-B., {Sun}, H., {et~al.} 2023, \apjl, 954, L17,
  \dodoi{10.3847/2041-8213/acf04a}

\bibitem[{{Yin} {et~al.}(2024){Yin}, {Zhang}, {Yang}, {Sun}, {Zhang}, {Shao},
  {Hu}, {Zhu}, {Xu}, {An}, {Gao}, {Wu}, {Zhang}, {Castro-Tirado}, {Pandey},
  {Rau}, {Lei}, {Xie}, {Ghirlanda}, {Piro}, {O'Brien}, {Troja}, {Jonker}, {Yu},
  {An}, {Chen}, {Chen}, {Dong}, {Eyles-Ferris}, {Fan}, {Fu}, {Fynbo}, {Gao},
  {Huang}, {Jiang}, {Jiang}, {Julakanti}, {Kuulkers}, {Lao}, {Li}, {Ling},
  {Liu}, {Liu}, {Mou}, {Pan}, {Wei}, {Wu}, {Yadav}, {Yang}, {Yuan}, \&
  {Zhang}}]{2024ApJ...975L..27Y}
{Yin}, Y.-H.~I., {Zhang}, B.-B., {Yang}, J., {et~al.} 2024, \apjl, 975, L27,
  \dodoi{10.3847/2041-8213/ad8652}

\bibitem[{{Yin} {et~al.}(2025){Yin}, {Fang}, {Zhang}, {Deng}, {Yang}, {Chen},
  {Liu}, {Cheng}, {Xu}, {Wang}, {Shen}, {Li}, {Mao}, {Li}, {Castro-Tirado},
  {Lei}, {Fu}, {Yang}, {Jiang}, {An}, {Chen}, {Dong}, {Du}, {Esamdin}, {Fan},
  {Feng}, {Feng}, {Fern{\'a}ndez-Garc{\'\i}a}, {Gao}, {Gritsevich}, {Guo},
  {Hu}, {Hu}, {Hua}, {Iskandar}, {Jin}, {Li}, {Li}, {Li}, {Lin}, {Liu}, {Liu},
  {Liu}, {Liu}, {Liu}, {Malesani}, {P{\'e}rez-Garc{\'\i}a}, {Sun}, {Wu},
  {Xiao}, {Xiong}, {Yan}, {Zhang}, {Zhang}, {Zhu}, {Zhu}, {Zou}, {Yuan}, \&
  {Zhang}}]{2025arXiv250600435Y}
{Yin}, Y.-H.~I., {Fang}, Y., {Zhang}, B.-B., {et~al.} 2025, arXiv e-prints,
  arXiv:2506.00435, \dodoi{10.48550/arXiv.2506.00435}

\bibitem[{{Yuan} {et~al.}(2025){Yuan}, {Dai}, {Feng}, {Jin}, {Jonker},
  {Kuulkers}, {Liu}, {Nandra}, {O'Brien}, {Piro}, {Rau}, {Rea}, {Sanders},
  {Tao}, {Wang}, {Wu}, {Zhang}, {Zhang}, {Ai}, {Buchner}, {Bulbul}, {Chen},
  {Chen}, {Chen}, {Chen}, {Coleiro}, {Coti Zelati}, {Dai}, {Fan}, {Fan},
  {Friedrich}, {Gao}, {Ge}, {Ge}, {Geng}, {Ghirlanda}, {Gianfagna}, {Gou},
  {Guillot}, {Hou}, {Hu}, {Huang}, {Ji}, {Jia}, {Komossa}, {Kong}, {Lan}, {Li},
  {Li}, {Li}, {Li}, {Li}, {Li}, {Ling}, {Liu}, {Liu}, {Liu}, {Liu}, {Luo},
  {Ma}, {Maggi}, {Maitra}, {Marino}, {Chi-Yung Ng}, {Pan}, {Rukdee}, {Soria},
  {Sun}, {Tam}, {Linesh Thakur}, {Tian}, {Troja}, {Wang}, {Wang}, {Wang},
  {Wei}, {Wen}, {Wu}, {Wu}, {Xiao}, {Xu}, {Xu}, {Xu}, {Xu}, {Yang}, {You},
  {Yu}, {Yu}, {Zhang}, {Zhang}, {Zhang}, {Zhang}, {Zhang}, {Zhang}, {Zhou}, \&
  {Zou}}]{2025arXiv250107362Y}
{Yuan}, W., {Dai}, L., {Feng}, H., {et~al.} 2025, arXiv e-prints,
  arXiv:2501.07362, \dodoi{10.48550/arXiv.2501.07362}

\bibitem[{{Zhang}(2006)}]{2006Natur.444.1010Z}
{Zhang}, B. 2006, \nat, 444, 1010, \dodoi{10.1038/4441010a}

\bibitem[{{Zhang} {et~al.}(2009){Zhang}, {Zhang}, {Virgili}, {Liang}, {Kann},
  {Wu}, {Proga}, {Lv}, {Toma}, {M{\'e}sz{\'a}ros}, {Burrows}, {Roming}, \&
  {Gehrels}}]{2009ApJ...703.1696Z}
{Zhang}, B., {Zhang}, B.-B., {Virgili}, F.~J., {et~al.} 2009, \apj, 703, 1696,
  \dodoi{10.1088/0004-637X/703/2/1696}

\bibitem[{{Zhang} {et~al.}(2023){Zhang}, {Huang}, {Liu}, \&
  {Wang}}]{2023ApJ...956L..21Z}
{Zhang}, H.-M., {Huang}, Y.-Y., {Liu}, R.-Y., \& {Wang}, X.-Y. 2023, \apjl,
  956, L21, \dodoi{10.3847/2041-8213/acfcab}

\bibitem[{{Zhou} {et~al.}(2024){Zhou}, {Chen}, {Sun}, {Zhang}, {Hu}, {Li},
  {Ling}, {Liu}, {Zhang}, {Jin}, {Cheng}, {Cui}, {Fan}, {Hu}, {Huang}, {Liu},
  {Liu}, {Lv}, {Lian}, {Mao}, {Pan}, {Pan}, {Wang}, {Wang}, {Wu}, {Xu}, {Xu},
  {Yang}, {Yuan}, {Zhang}, {Zhang}, {Zhang}, {Zhang}, {Zhao}, {Yang}, {Dai},
  {Liang}, {Chen}, {Jia}, {Zhang}, {Kuulkers}, {Santovincenzo}, {O'Brien},
  {Nandra}, {Rau}, {Cordier}, \& {Einstein Probe Team}}]{GCN36691}
{Zhou}, H., {Chen}, W., {Sun}, H., {et~al.} 2024, GRB Coordinates Network,
  36691, 1

\end{thebibliography}
\bibliographystyle{aasjournal}

\appendix\setcounter{figure}{0}
\renewcommand{\thefigure}{A\arabic{figure}}
\setcounter{table}{0}
\renewcommand{\thetable}{A\arabic{table}}

\begin{figure*}[!htp]
    \centering
    \includegraphics[width=0.49\textwidth]{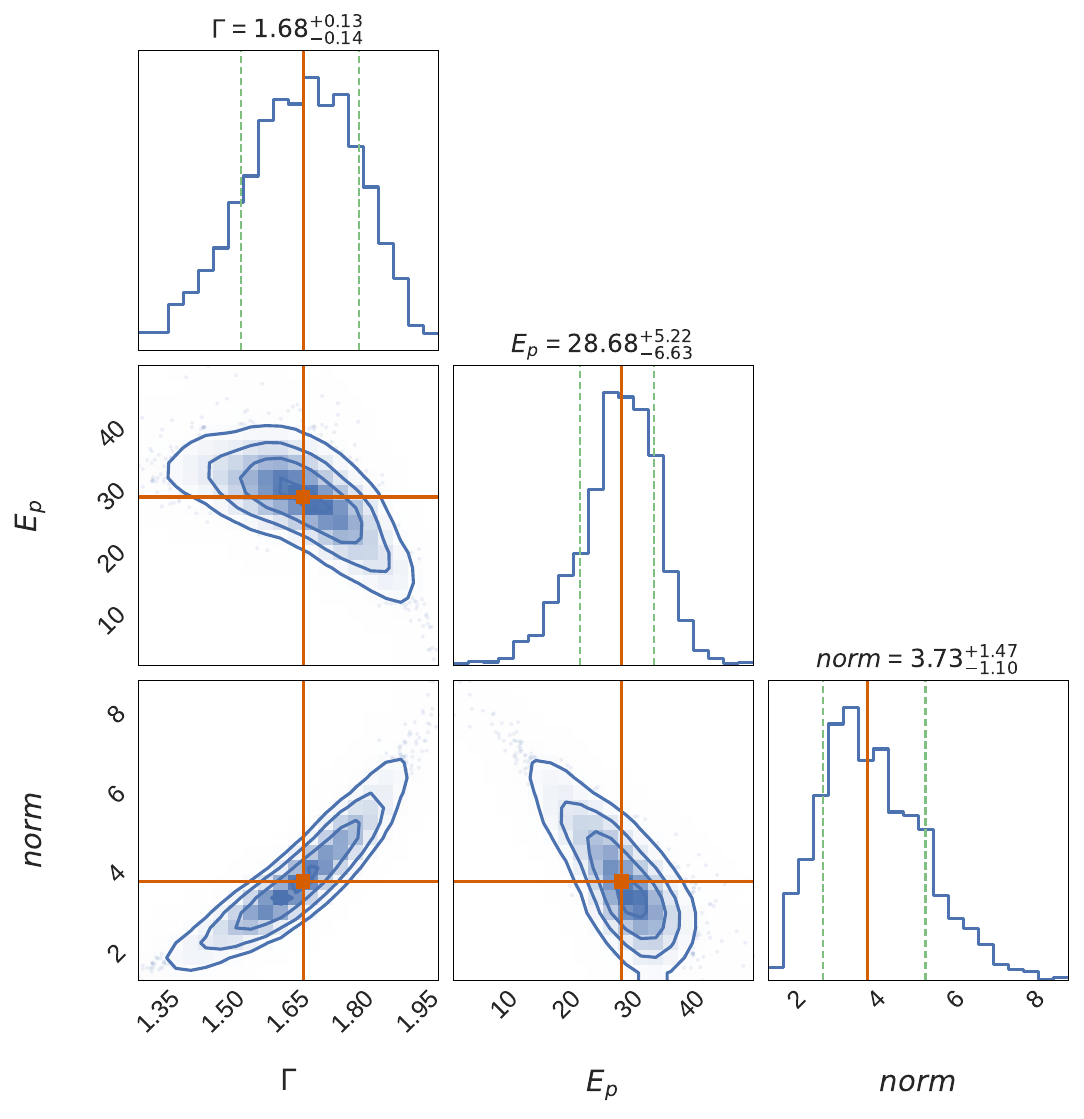}
    \includegraphics[width=0.49\textwidth]{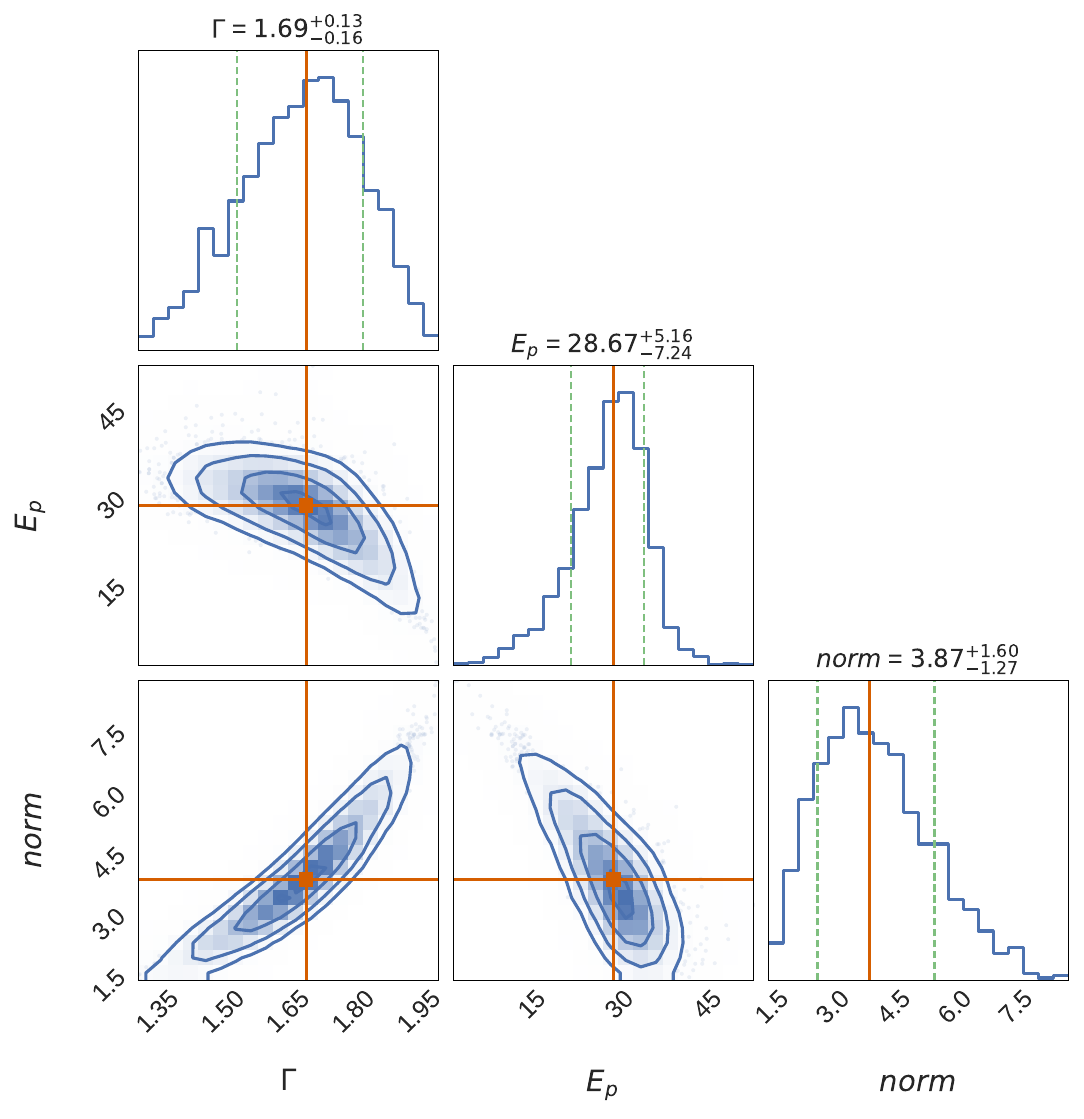}
    \caption{The posterior distributions from fitting the CPL model using response matrices generated from historical pointing files and real-time pointing files, corresponding to the left and right panels, respectively. The red line represents the median value, while the green dashed lines indicate the 1$\sigma$ confidence interval.}
    \label{fig:spec_compare}
\end{figure*}

In the Appendix, we present the spectral fitting results for different time intervals and instruments in Table~\ref{tab:specfit}, and a comparison of the posterior parameter distributions derived using response matrices generated from both historical and real-time pointing files in Figure~\ref{fig:spec_compare}.

\begin{table*}[htbp]
\centering
\begin{threeparttable}
\caption{Spectral fitting results}
\label{tab:specfit}
\begin{tabular}{cccccccc}
\hline
Instrument & Time Interval & \multicolumn{3}{c}{$tbabs$*(PL/CPL) \tnote{a}} & $\ln{\cal Z}$ & Flux$_{\rm 0.5-4 keV}$ & Flux$_{\rm 8-900 keV}$\\
\cline{3-5}
& [s] & $\Gamma$ & $E_{\rm p}$ [keV] & $N_{\rm H}$ [$10^{22}$\,cm$^{-2}$] & & \multicolumn{2}{c}{10$^{-8}$ [erg\,cm$^{-2}$\,s$^{-1}$]} \\
\hline
WXT & a: (-33, 217) & $1.87_{-0.40}^{+0.34}$ & ... & $0.39_{-0.13}^{+0.13}$ & -14.70 & $0.18_{-0.02}^{+0.03}$  &...\\
WXT & b: (217, 289) & $1.83_{-0.23}^{+0.22}$ & ... & $0.47_{-0.08}^{+0.09}$& -16.49 & $1.44_{-0.12}^{+0.15}$ &...\\
WXT & tot: (-33, 289) & $1.67_{-0.15}^{+0.16}$ & ... & $0.37_{-0.05}^{+0.06}$ & -34.16 &$0.43_{-0.02}^{+0.03}$ &...  \\
\hline
WXT+GBM(nb) & b: (217, 289) & $2.06_{-0.03}^{+0.02}$ & ... & $0.57_{-0.04}^{+0.05}$ & -207.59 & $1.63_{-0.11}^{+0.11}$ & $2.93_{-0.15}^{+0.16}$\\
WXT+GBM(nb) & b: (217, 289) & $1.61_{-0.05}^{+0.05}$ & $33.31_{-3.87}^{+5.06}$ & $0.40_{-0.05}^{+0.05}$ & -200.36 & $1.37_{-0.10}^{+0.10}$ & $4.19_{-0.25}^{+0.26}$ \\
\hline
GBM(nb) & $T_{90}$: (227, 347) & $1.68_{-0.14}^{+0.13}$ & $28.68_{-6.63}^{+5.22}$ & ... & -268.88 & ...& $3.73_{-1.10}^{+1.47}$\\
GBM(nb)\tnote{b} & $T_{90}$: (227, 347) & $1.69_{-0.16}^{+0.13}$ & $28.67_{-7.24}^{+5.16}$ & ... & -268.35 & ... & $3.87_{-1.27}^{+1.60}$ \\
\hline
\end{tabular}
\begin{tablenotes}
\footnotesize
\item[a] When $E_{\rm p}$ is denoted by ``...'', it indicates that the model used is PL. The listed values represent the median values and the corresponding 68\% credible intervals of the posterior distribution. When $N_{\rm H}$ is denoted by ``...'', it indicates that the absorption model $tbabs$ was not considered in the analysis.
\item[b] The response matrices used in the fitting were generated using the historical pointing files.
\end{tablenotes}
\end{threeparttable}
\end{table*}


\end{document}